\documentclass[%
 aip,
jap,
numerical,
 amsmath,amssymb,
]{revtex4-1}

\usepackage{graphicx}
\usepackage{dcolumn}
\usepackage{bm}
\usepackage[export]{adjustbox}
\graphicspath{ {Figures/} }
\usepackage{fnpct}
\usepackage{hyperref}
\usepackage{amsmath}
\usepackage{amsxtra}
\usepackage{amstext}
\usepackage{amssymb}
\usepackage{latexsym}
\usepackage{dsfont}
\usepackage[capitalise]{cleveref}
\usepackage{textcomp}
\usepackage{ulem}
\usepackage{siunitx}
\usepackage{multirow}
\usepackage{color}
\usepackage{caption, subcaption}
\usepackage{soul}
\usepackage{natbib}
\usepackage{mathtools,upgreek}
\usepackage{tabularx,ragged2e,booktabs,caption}
\usepackage{bigdelim,bigstrut}
\newcolumntype{C}[1]{>{\Centering}m{#1}}

\renewcommand{\vec}[1]{\bm{\mathrm{#1}}}
\newcommand{\etal}{\textit{et~al}. }
\hypersetup{
	colorlinks=true,
	citecolor=blue,
	linkcolor=blue,
	filecolor=blue,      
	urlcolor=blue,
}

\begin{document}


\title{Reduced order derivation of the two-dimensional band structure of a mixed-mode resonator array}

\author{Alireza V. Amirkhizi}
 \email{alireza\_amirkhizi@uml.edu}
\author{Weidi Wang}
\noaffiliation
\affiliation{Department of Mechanical Engineering, University of Massachusetts, Lowell, 
Lowell, Massachusetts 01854, USA
}%

\date{\today}

\begin{abstract}
In this paper, the 2D band structure of a mixed-mode metamaterial resonator array for in-plane waves is investigated. The band structure in the interior and on the boundary of the irreducible Brillouin zone as well as 1D dispersion diagrams for different propagation angles are calculated numerically and presented. Additionally, a reduced order analytical method is established to compare and approximate the band structure. The studied metamaterial, with a T-shaped cantilever beam as resonator in its square array repeating unit cells, exhibits branches with mixed P and SV waves except at exactly one angle of propagation. This paper also reports on the occurrence of avoided level crossings, which are related to the existence of exceptional points in the complex domain. A reduced order analytical approach is used that can generate partial (low branches) band structure with relatively little computational effort. The reduced order model agrees well with the numerical results for these low branches and can provide support in mode identification and band sorting. With proper adjustments in parameters, this analytical method will be applicable for other metamaterials that have similar unit cell structure. 
\end{abstract}

\maketitle

\section{Introduction}

Acoustic metamaterials and phononic crystals\cite{Liu2000} have attracted a lot of attention in the past two decades mostly because of their potential ability to manipulate waves and their unusual overall dynamic properties. These novel dynamic properties lead to many proposed applications, such as wave filtering \cite{Sun2006,Sun2007}, attenuation\cite{Kushwaha1994,Nemat-Nasser2015}, negative refractive index\cite{Ding2010,Seo2012,Li2016}, and cloaking\cite{Chen2007,Norris2011}.

Due to the local resonance in their periodic structure, such media exhibit frequency band gaps, within which there are no propagating waves. The determination of band structure is of prime importance as the natural first step in deriving the overall properties of metamaterials. In particular the overall dynamic properties are generally close to what one might estimate using quasi-static micromechanical methods, except near resonances and band gaps. 

The material properties and geometrical layouts are the main factors affecting the band structure. 
Claeys \etal\cite{Claeys2017} investigated various metamaterial layouts for reducing vibrations along a known transmission path both numerically and experimentally. Wang \etal\cite{Wang2011} studied the influence of geometry on band gap properties of phononic crystals and developed spring-mass models of different vibrational modes to predict band gap frequencies. Even though most applications are meant for control of acoustic waves, the solid structure can inherently maintain shear and mixed modes. The present work is particularly interested in understanding the nature of such mixed modes in mechanical metamaterials. 

Various methods have been developed for band structure calculation. Transfer Matrix (TM) method is widely used\cite{Yu2006a,Wang2013b,Nemat-Nasser2015,Sridhar2017,Zuo2017,Amirkhizi2017,Nanda2018} in 1D periodic beams and multi-layered structure. However, the TM method usually can only be applied in one direction. Plane Wave Expansion (PWE) method, which expands the material property and amplitude of Bloch wave response into Fourier series, is one of the most commonly used approaches\cite{Mead1996,Wu2004b,Xiao2012,Oudich2014,Zhu2014a} for calculating the dispersion relations of periodic structures including both phononic and photonic crystals. The PWE method is easy to apply and has wide applications, but it has slow convergence rate for systems of large elastic mismatch.
Nemat-Nasser \etal\cite{Nemat-Nasser2011} introduced a mixed variational method which has high accuracy and quick convergence for calculating dispersion relation and effective elastodynamic parameters of periodic elastic composites. For phononic crystals or metamaterials with complex microstructures, finite element method (FEM) provides efficient and precise numerical solutions.

Beams are commonly found in locally resonant media as they could provide a compact ``spring'' element.
Wang and Wang \cite{Wang2013} developed spring-mass models for three-dimensional phononic crystals and evaluated the band gap edge frequencies based on effective stiffness of beams. On the other hand, as the main medium for 1D propagation with attached resonators, Timoshenko beam theory has been used by Yu \etal\cite{Yu2006a,Yu2006} to study the band structures of metamaterial beams with the transfer matrix method. Extensive studies\cite{Xiao2012,Xiao2013,Wang2013b,Zuo2017,Hu2018} have been done on locally resonant structures based on Euler-Bernoulli and Timoshenko beam theories.

Wu \etal\cite{Wu2004b} studied band gaps of surface waves in two-dimensional phononic structures consisting of general anisotropic materials based on PWE and reported that some of the apparent crossing points in dispersion curves are indeed sharp bends around which the modes exchange suddenly. What appears incorrectly as a crossing point could lead to incorrect sorting of the band structure.
Such level repulsion in dispersion curves has been studied and reported in literatures. Wu and Huang \cite{Wu2004a} analyzed level repulsions of bulk acoustic waves and pointed out that the polarizations of different modes could be used as the criterion to identify real or apparent crossing points.
Numerical and experimental studies performed by Yeh \etal\cite{Yeh2016} showed that the band gap induced by level repulsion strongly depend on the geometry of phononic structure. Level repulsions occur in the vicinity of exceptional points\cite{heiss1990} in complex wavenumber domain. Maznev\cite{Maznev2018a} analytically studied the complex band structure and eigenvectors of locally resonant media in the vicinity of exceptional point and discussed the effect of damping ratio.
Lu and Srivastava\cite{Lu2018} introduced a method based on the mode shape continuity around the exceptional point to identify the crossing points in the band structure, distinguish them from level repulsion or avoided crossings, and sort the bands correctly.  

\begin{figure}[!ht]
\begin{subfigure}[b]{0.5\linewidth}
\centering\includegraphics[height=150pt]{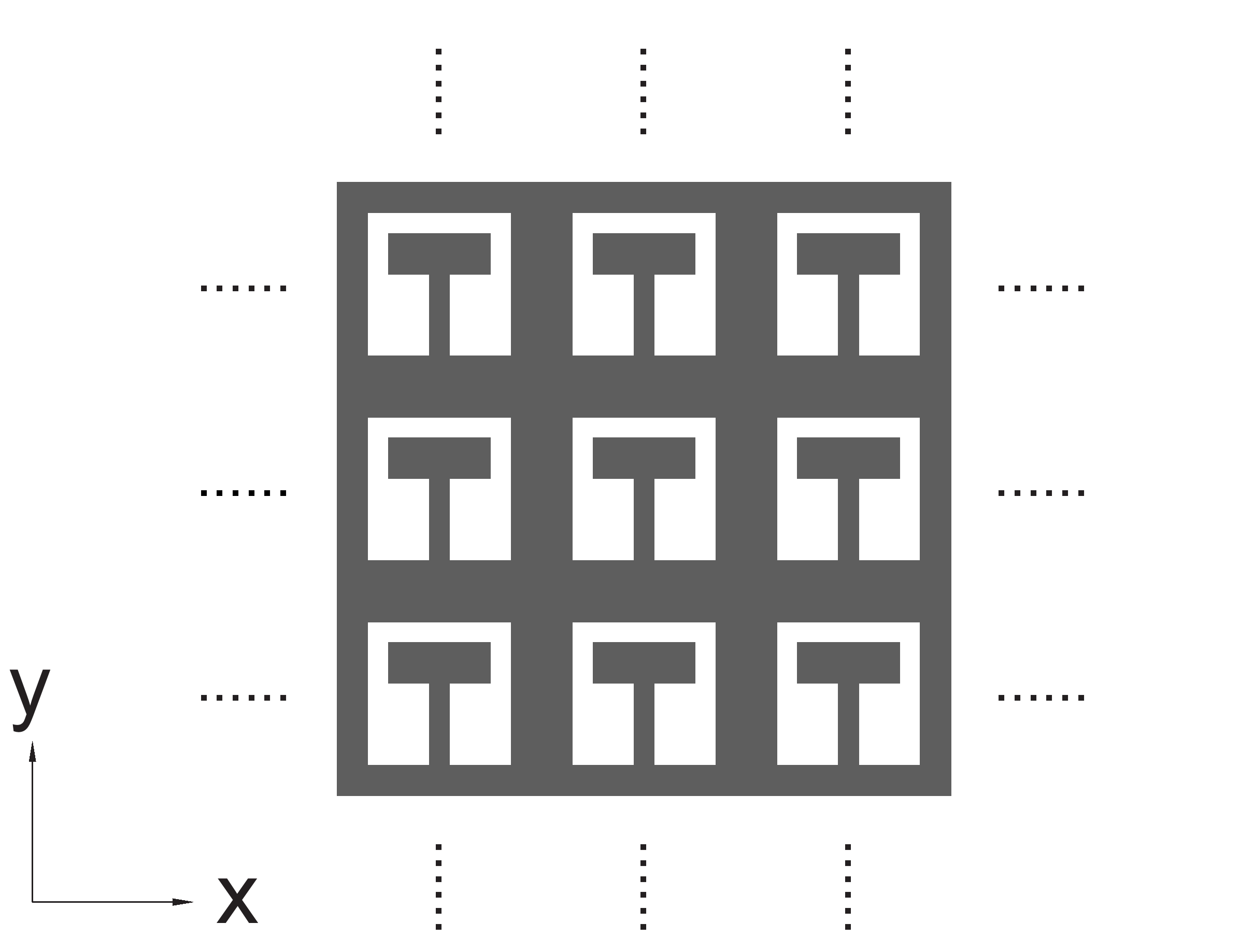}
\caption{\label{fig:schematic1}}
\end{subfigure}%
\begin{subfigure}[b]{0.5\linewidth}
\centering\includegraphics[height=130pt]{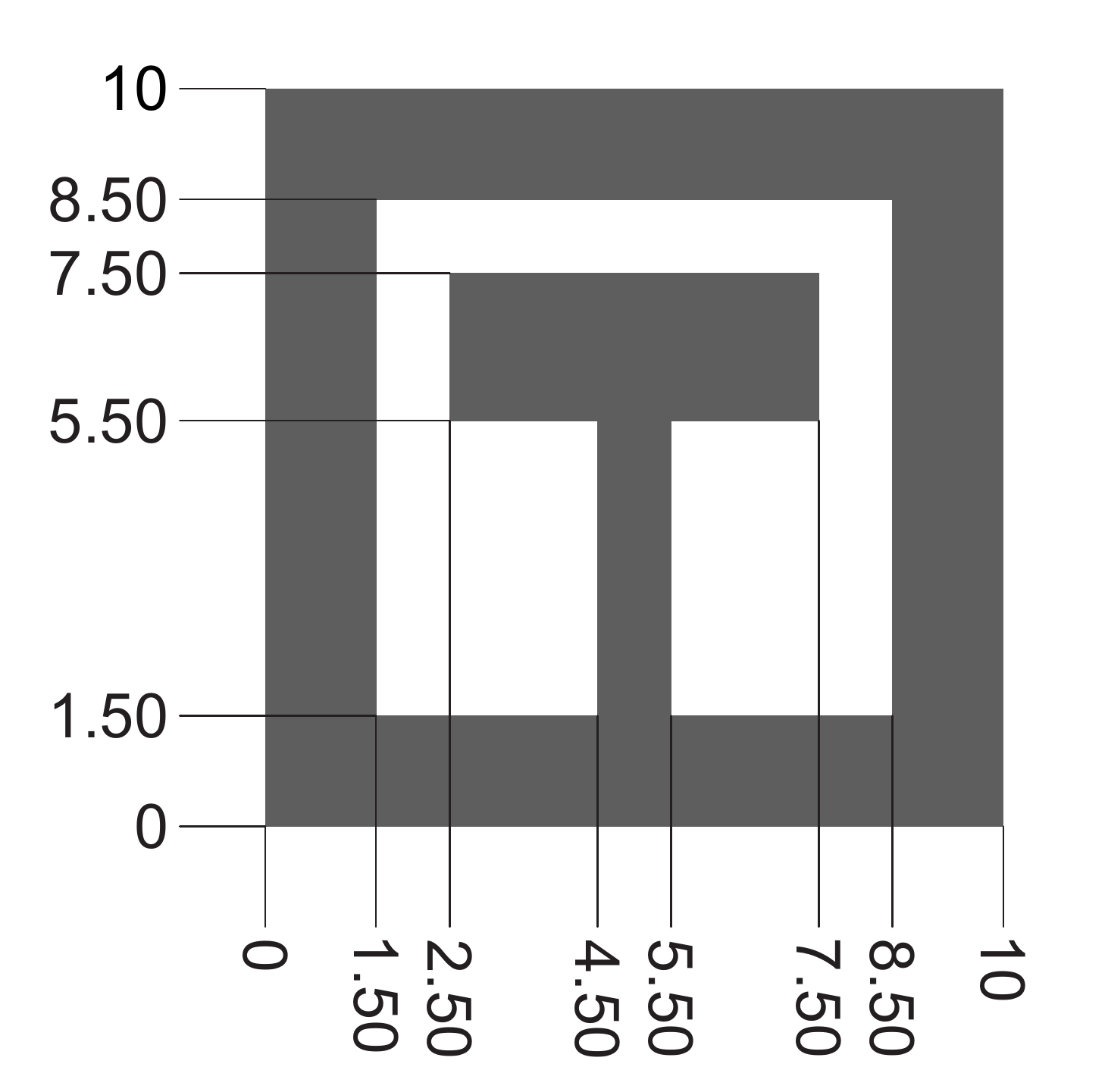}
\caption{\label{fig:ruc1}}
\end{subfigure}
\caption{(\subref{fig:schematic1}) Schematic drawing of the studied 2D infinitely periodic resonator array. (\subref{fig:ruc1})~The detailed geometry of the RUC with coordinates of each line shown in \si{mm}. \label{fig:structure}} 
\end{figure}

In this paper, the band structure of a resonator array for any wave with oblique in-plane propagation direction is investigated. The schematic of repeating unit cell (RUC) of the studied metamaterial is shown in \cref{fig:structure}. The RUC includes a T-shaped cantilever beam as the resonator and the hollow frame as the cell. It is further assumed that the system is in plane strain conditions with the thickness of the array much larger than cell and wall length scales. Such structure is easy to produce by 3D printing and can be analyzed as an assemblage of elastic beams and masses. It is expected to have local resonance at around \SI{3}{\kilo \hertz} based on preliminary estimates of the resonator. The incident angles of P and SV waves are varied from 0$^\circ$ ($x$ direction) to 90$^\circ$ ($y$ direction). The propagating P and SV waves are assumed to be plane harmonic Bloch waves. In what follows, we first present the band structure results obtained from numerical approach (FEM), including the dispersion diagrams and mode shapes. Next we establish an analytical method for band structure calculation based on identifying the major reduced order degrees of freedom (DOFs). Then we compare the results from two approaches and discuss how the dispersion curves change with respect to different propagation directions. Finally we present and discuss the occurrence of the avoided and real crossing points in the band structure.

\section{Numerical calculation of band structure}
\begin{figure}
\begin{subfigure}[c]{0.5\linewidth}
	\centering\includegraphics[height=145pt]{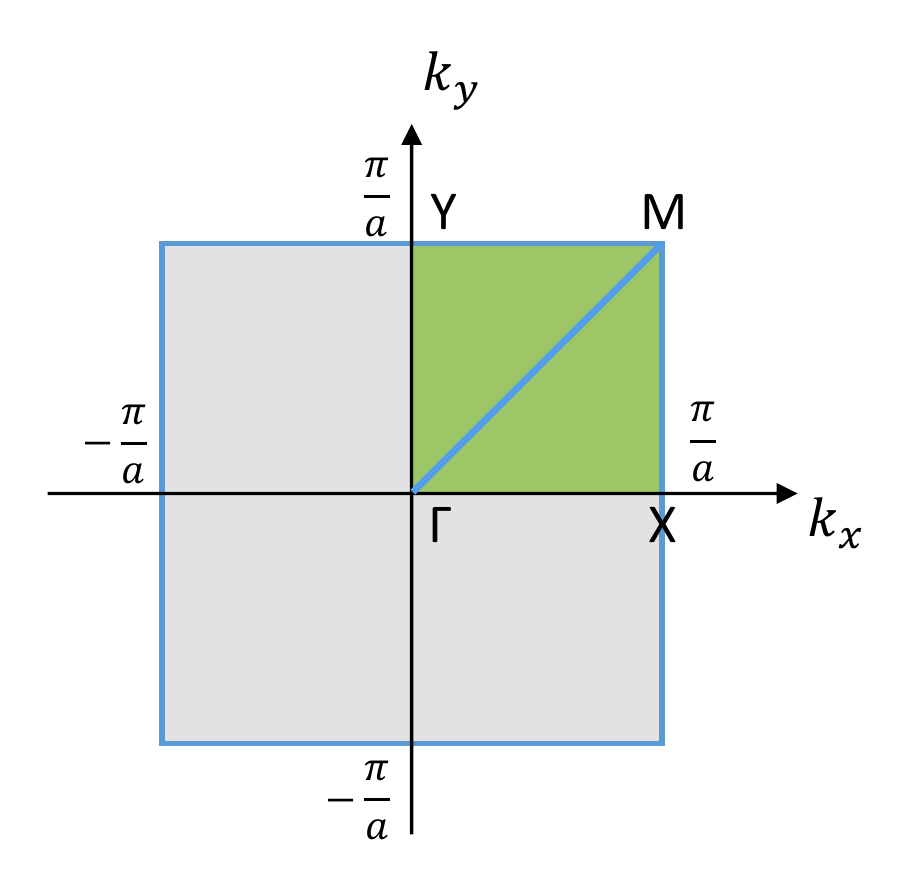}
	\caption{\label{fig:ibz}}
\end{subfigure}%
\begin{subfigure}[c]{0.5\linewidth}
	\centering\includegraphics[width=180pt]{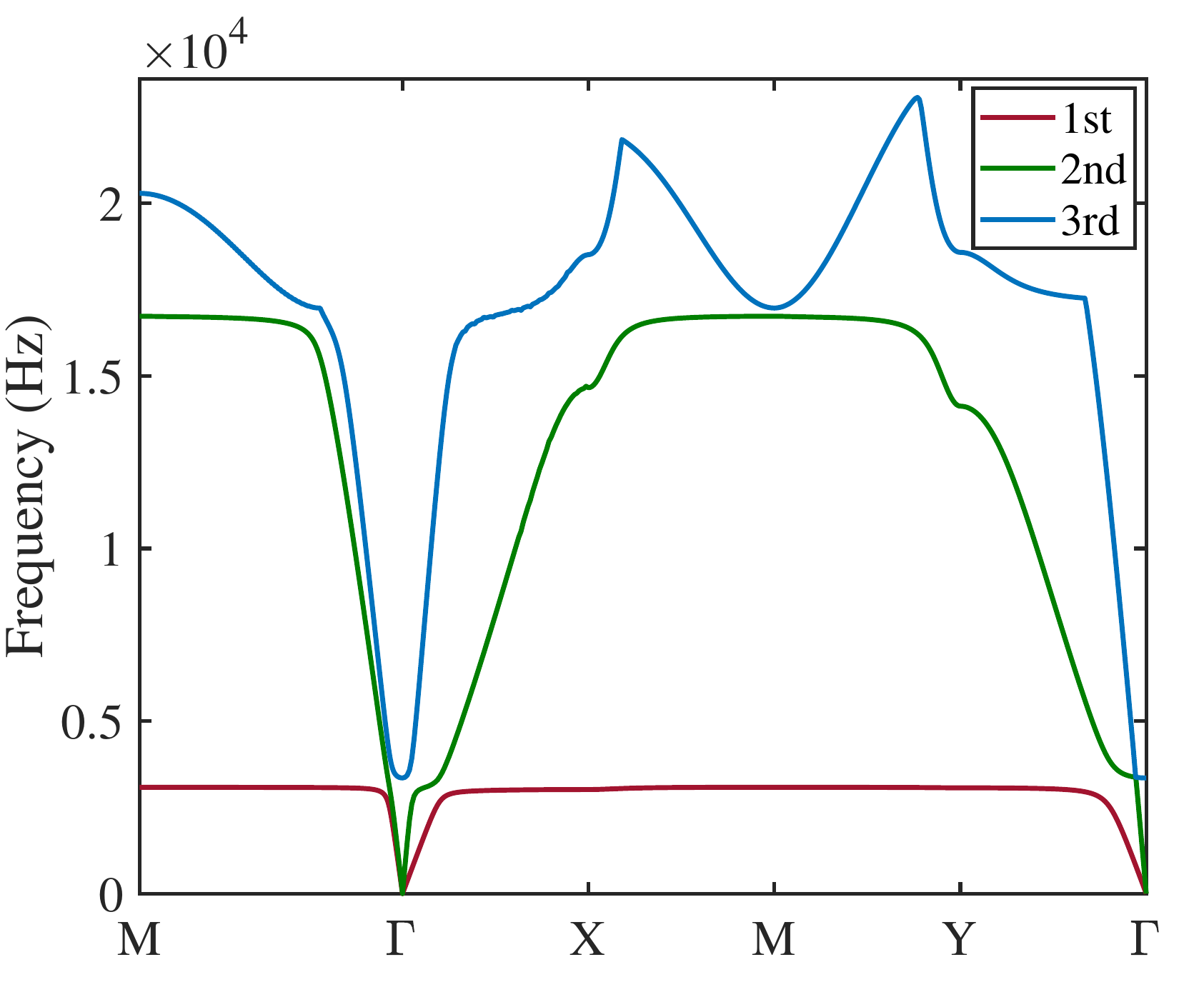}
	\caption{\label{fig:ibzband}}
\end{subfigure}
\caption{\label{fig:bzo} First Brillouin zone and complete dispersion curves of first three modes along paths connecting high symmetry $\vec{k}$ points.} 
\end{figure}

The numerical simulation is performed using finite element software COMSOL. The dynamic problem of elastic waves propagating in a periodic medium can be described formally as the eigenvalue problem
\begin{equation}
\left(\vec{K}-\omega^2\vec{M}\right)\vec{U}=0,
\end{equation}
where $\omega$ is the angular eigenfrequency, and the real part of $\vec{u}(x,y,t) = \vec{U}(x,y,\omega)\mathrm{exp}\left[-i \omega t\right]$ represents a harmonic displacement field at this frequency, and $\vec{K}$ and $\vec{M}$ are the stiffness and mass operators in Fourier domain, respectively. The spatial domain of this problem is reduced to a single unit cell based on Bloch-Floquet periodicity which requires the function $\vec{u_0}$ in
\begin{equation}
\vec{u}=\vec{u_0}(x,y,\omega) \mathrm{exp}\left[-i\left(\omega t - k_x n a - k_y m b\right)\right],
\end{equation} 
to be periodic. Here $n,m \in \mathbb{Z}$, and $a$ and $b$ are unit cell dimensions in $x$ and $y$ directions (in this case $a = b = $\SI{10}{mm}), respectively, and $\vec{k}$ is the wavevector. The eigenfrequencies can be obtained as functions of $\vec{k}$ by solving this system, in particular, by applying a discretized mesh and converting the operator equation above to a finite matrix over the unit cell domain. In the present study, the structure is modeled based on the material properties of VeroGray, a widely available 3D printing material (density $\rho = $ \SI{1161}{\kilogram m /s^3}, Young's modulus $E = $ \SI{2.978}{\giga\pascal}, and Poisson's ratio $\nu = $ \SI{0.35}{}). We define non-dimensionalized wavevector components $k_x a = Q_x$ and $k_y a = Q_y$ in $x$ and $y$ directions, respectively, with
\begin{align}
Q_x&=Q\cos\theta,\\
Q_y&=Q\sin\theta,
\end{align}
where $\theta$ is the propagation angle and $Q$ is a dimensionless parameter. The 2D band structure can be obtained by solving eigenfrequencies with parametric sweep of $Q_x$ and $Q_y$ from 0 to $\pi$. 
The first three mode frequencies calculated along paths connecting high symmetry $\vec{k}$ points in the Irreducible Brillouin Zone (IBZ) are shown in \cref{fig:bzo}.

\begin{figure}[!ht]
\begin{subfigure}[c]{0.5\linewidth}
	\centering\includegraphics[height=150pt]{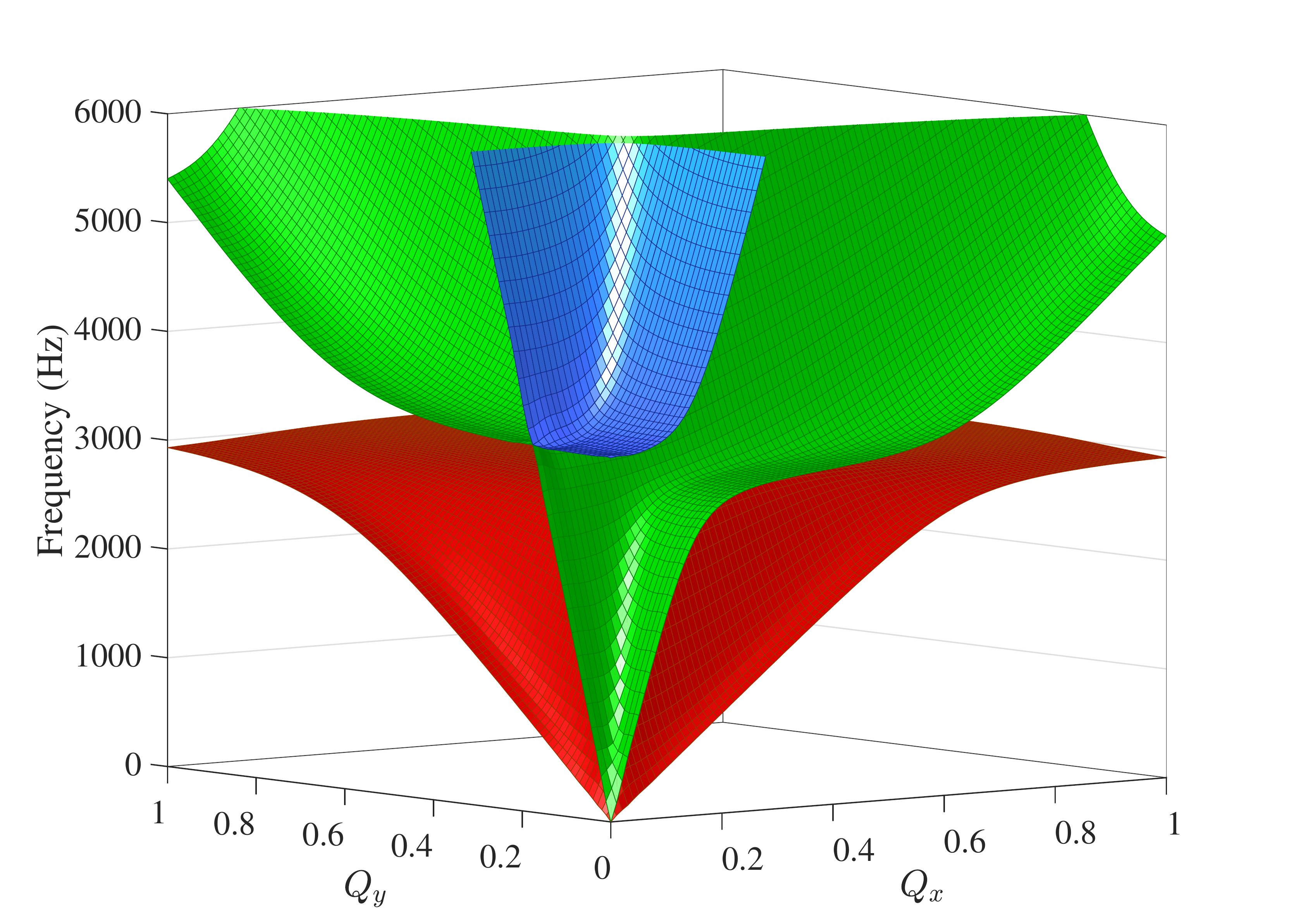}
	\caption{\label{fig:band3d}}
\end{subfigure}%
\begin{subfigure}[c]{0.5\linewidth}
\centering\includegraphics[width=150pt]{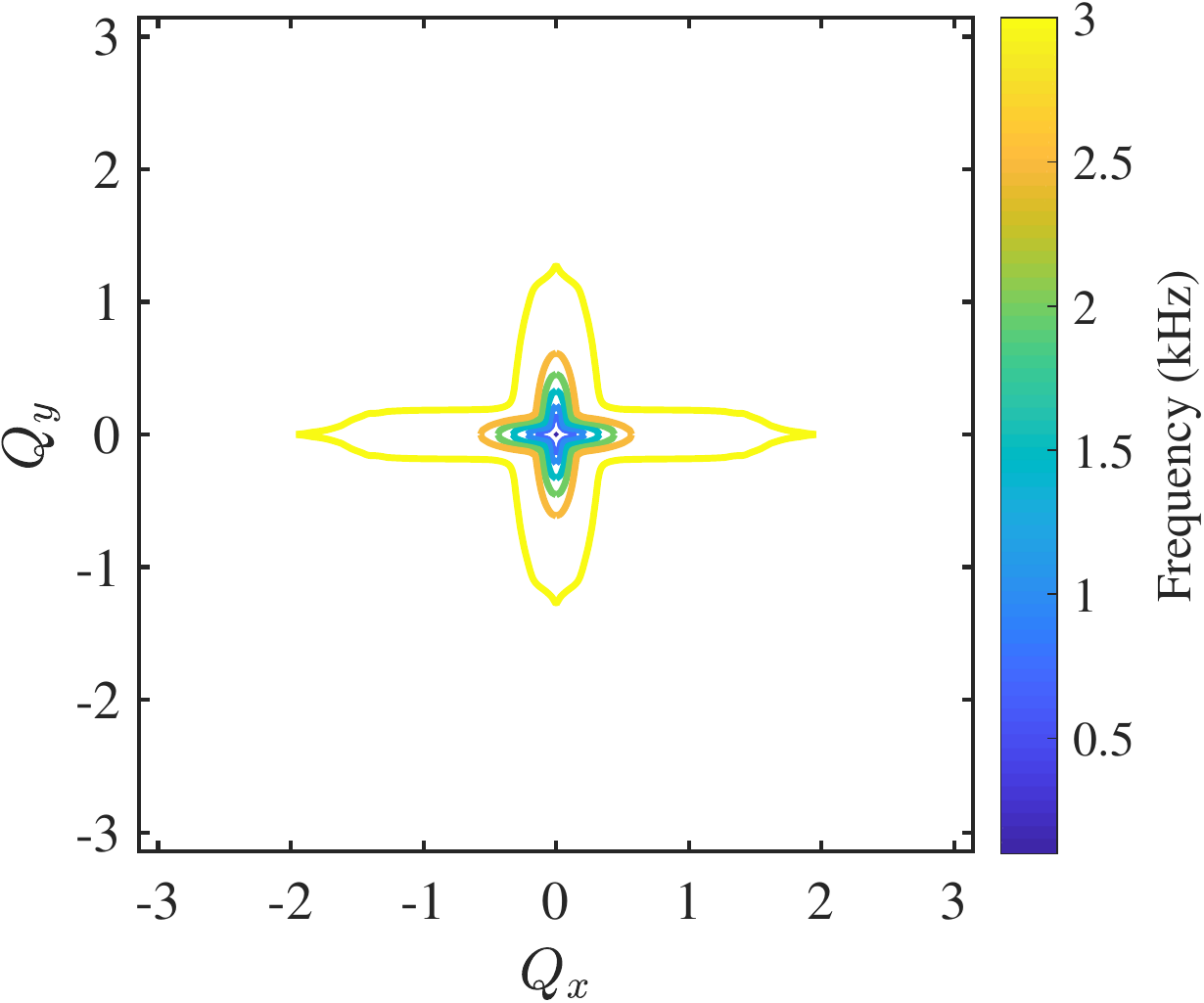}
\caption{\label{fig:cmode1}}
\end{subfigure}\\
\begin{subfigure}[c]{0.5\linewidth}
\centering\includegraphics[width=150pt]{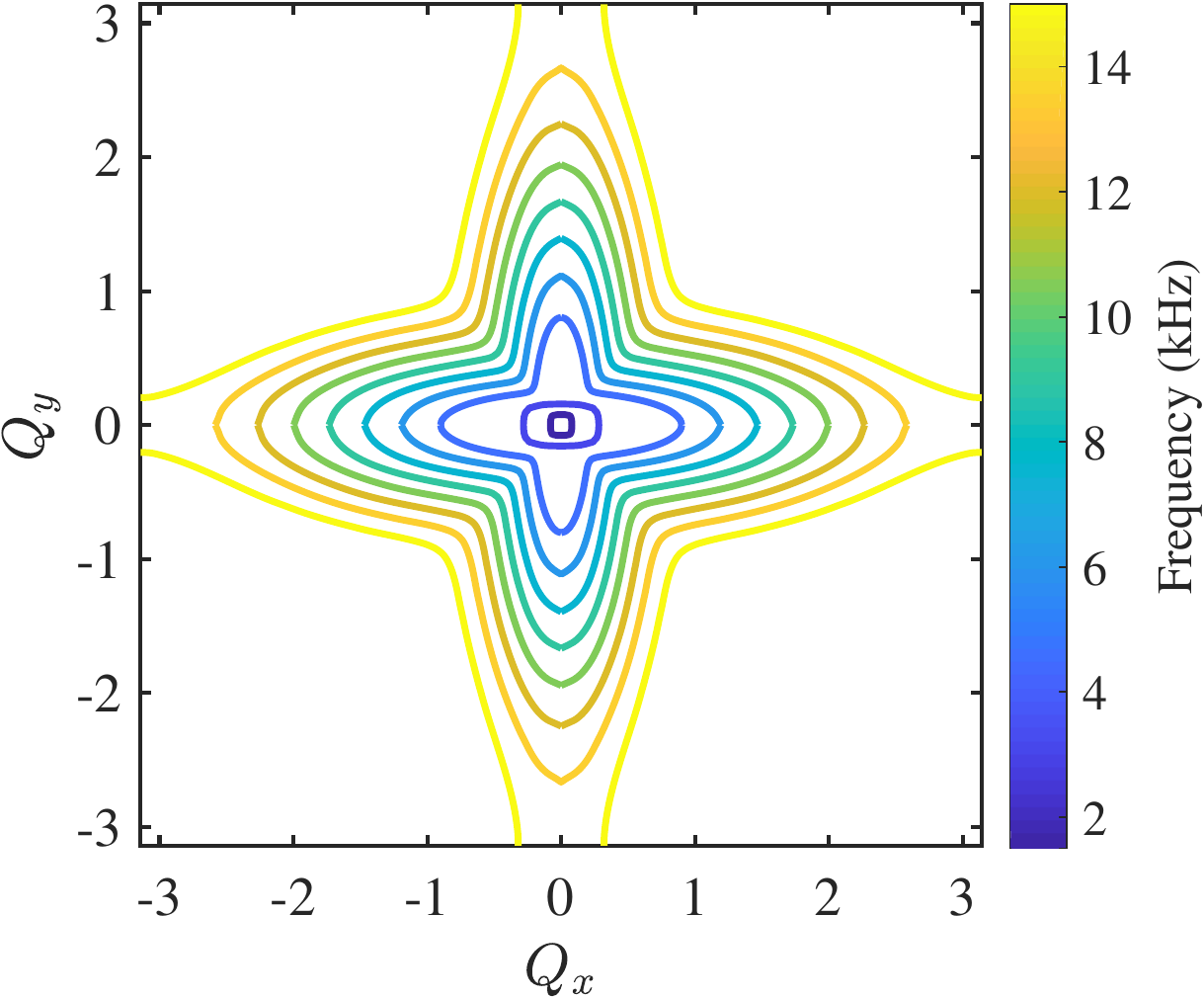}
\caption{\label{fig:cmode2}}
\end{subfigure}%
\begin{subfigure}[c]{0.5\linewidth}
\centering\includegraphics[width=150pt]{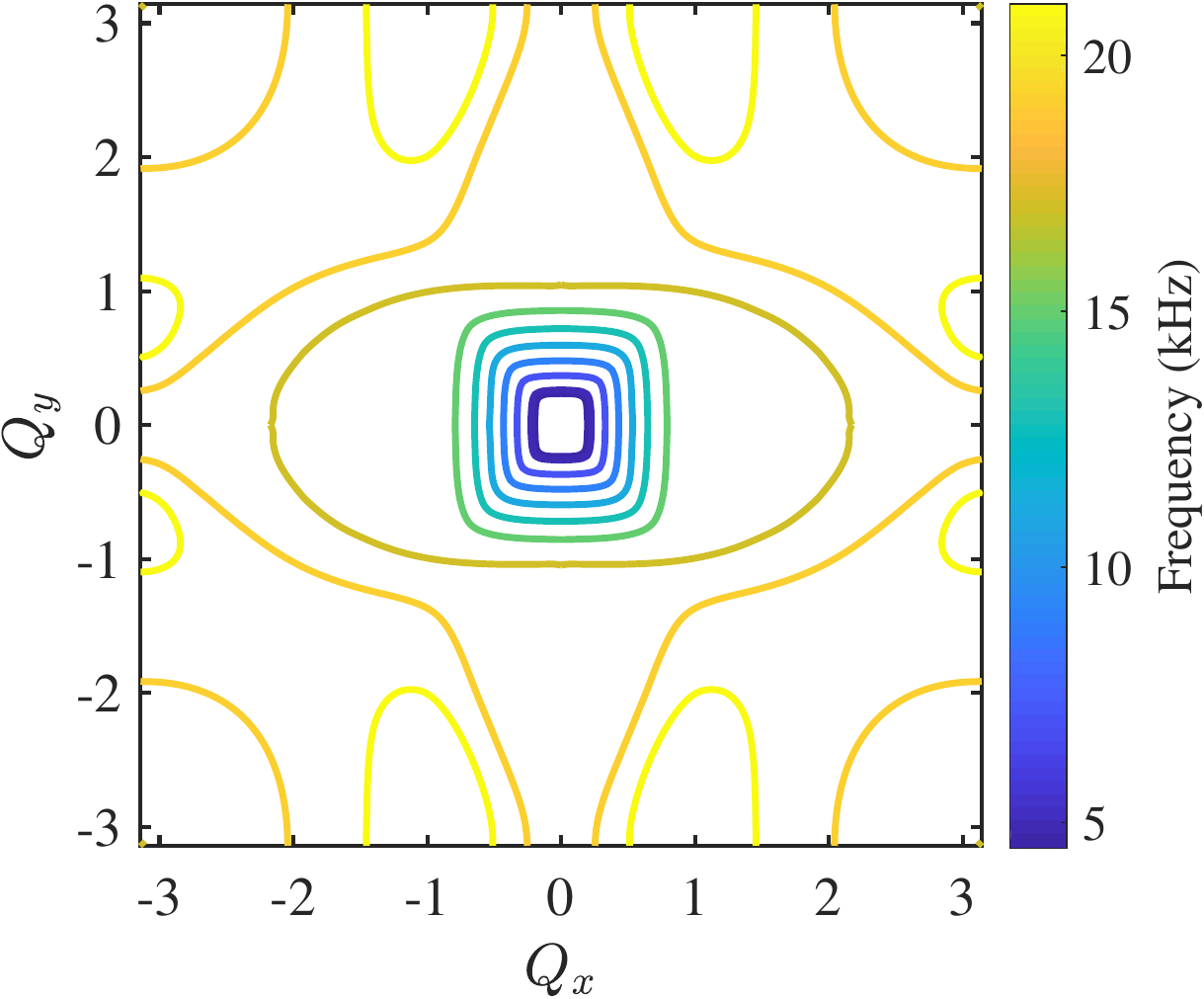}
\caption{\label{fig:cmode3}}
\end{subfigure}%
\caption{\label{fig:cont} Partial band structure showing first 3 modes of the studied metamaterial. Red, green, and blue surfaces in (\subref{fig:band3d}) represent the first, second, and third modes, respectively (color online). (\subref{fig:cmode1}) through (\subref{fig:cmode3}) show frequency contours of the first 3 modes, in order.} 
\end{figure}

The 3D partial band structure is shown in \cref{fig:band3d}. The frequency contour plots of the first three modes are shown in \crefrange{fig:cmode1}{fig:cmode3}, respectively. \Cref{fig:cmode1} show the frequency contour of the first mode (SV acoustic branch). Its first quadrant ($Q_x>0, Q_y>0$) is partially shown as the red surface in \cref{fig:band3d}. The contour curves a sharp drop in the phase around 45$^\circ$ angle of propagation. This is expected as one should note the cooperative response of the vertical and horizontal walls in bending leading to higher stiffness in the cell structure against in-plane shear at middle angles. Similar trend can also be seen in \cref{fig:cmode2}, where the contour curves for higher frequencies ($>3$ kHz) represent SV optical branch. The trend of P wave speed can be seen from the two inner curves in \cref{fig:cmode2}, representing its acoustic branch. Unlike SV wave, the P wave speed first decreases then increases as the propagation angle changes from 0$^\circ$ to 90$^\circ$. The underlying structural resistance to in-plane axial deformation changes from stretch-dominated near 0$^\circ$ and 90$^\circ$, to bending-dominated at middle angles, representing lower stiffness and phase speeds in contrast to shear response. \Cref{fig:cmode3} shows the contour of the third mode, which is mostly in a range beyond the considerations of this paper, but the central portion (lower frequencies) also match the P wave acoustic branch behavior and may be identified by its optical branch. Note that this representation provides more information compared to \cref{fig:bzo} as it potentially allows for inspection of interesting points in the reciprocal space not on the high symmetry paths. This is in particular important for the mechanical metamaterials as substantial mode mixing is inherent to the physics of the problem. 

\begin{figure}[!ht]
\begin{subfigure}[b]{0.5\linewidth}
\centering\includegraphics[height=175pt]{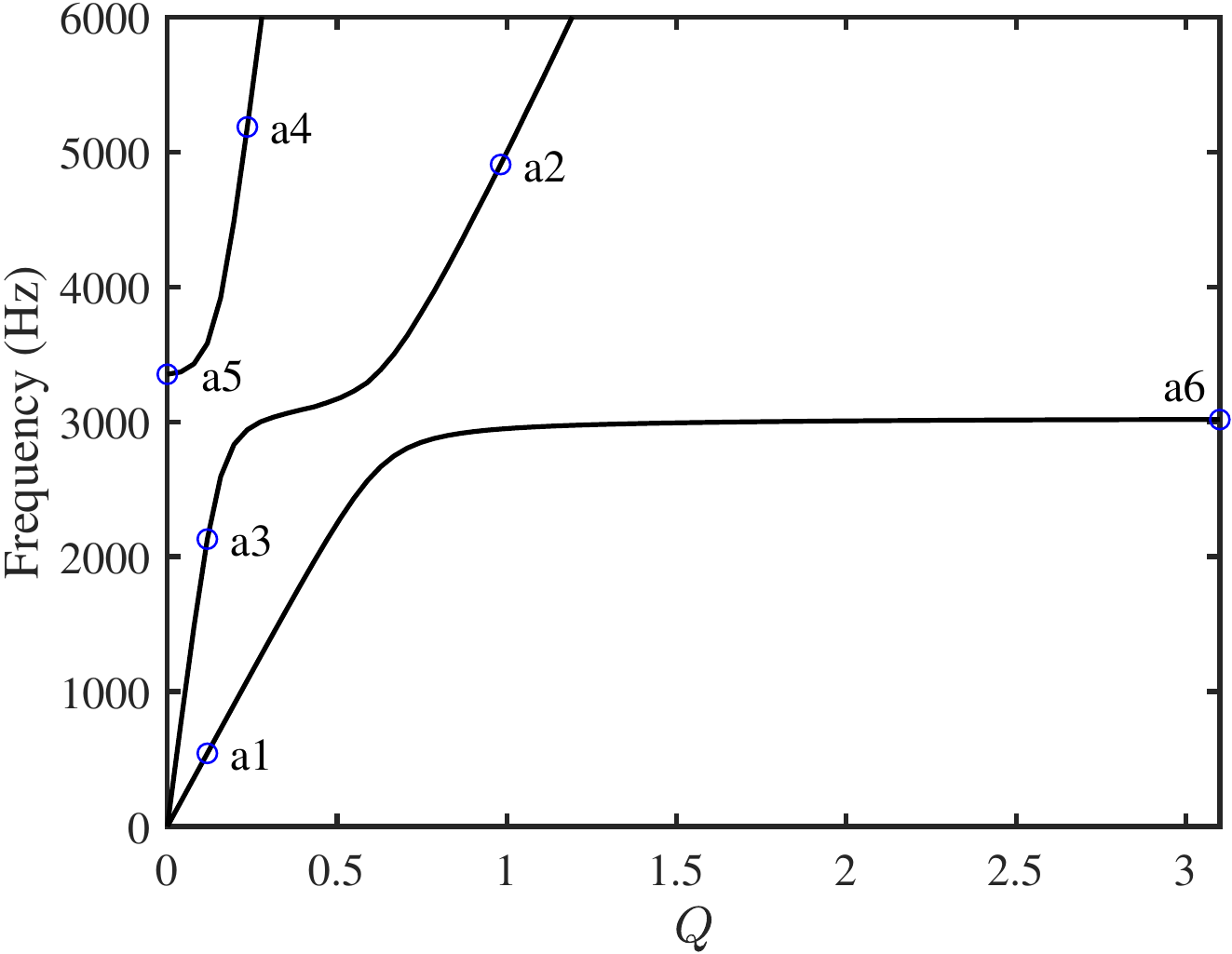}
\caption{$\theta = 0^\circ$ \label{fig:dc0comsoldots}}
\end{subfigure}%
\begin{subfigure}[b]{0.5\linewidth}
\centering\includegraphics[height=175pt]{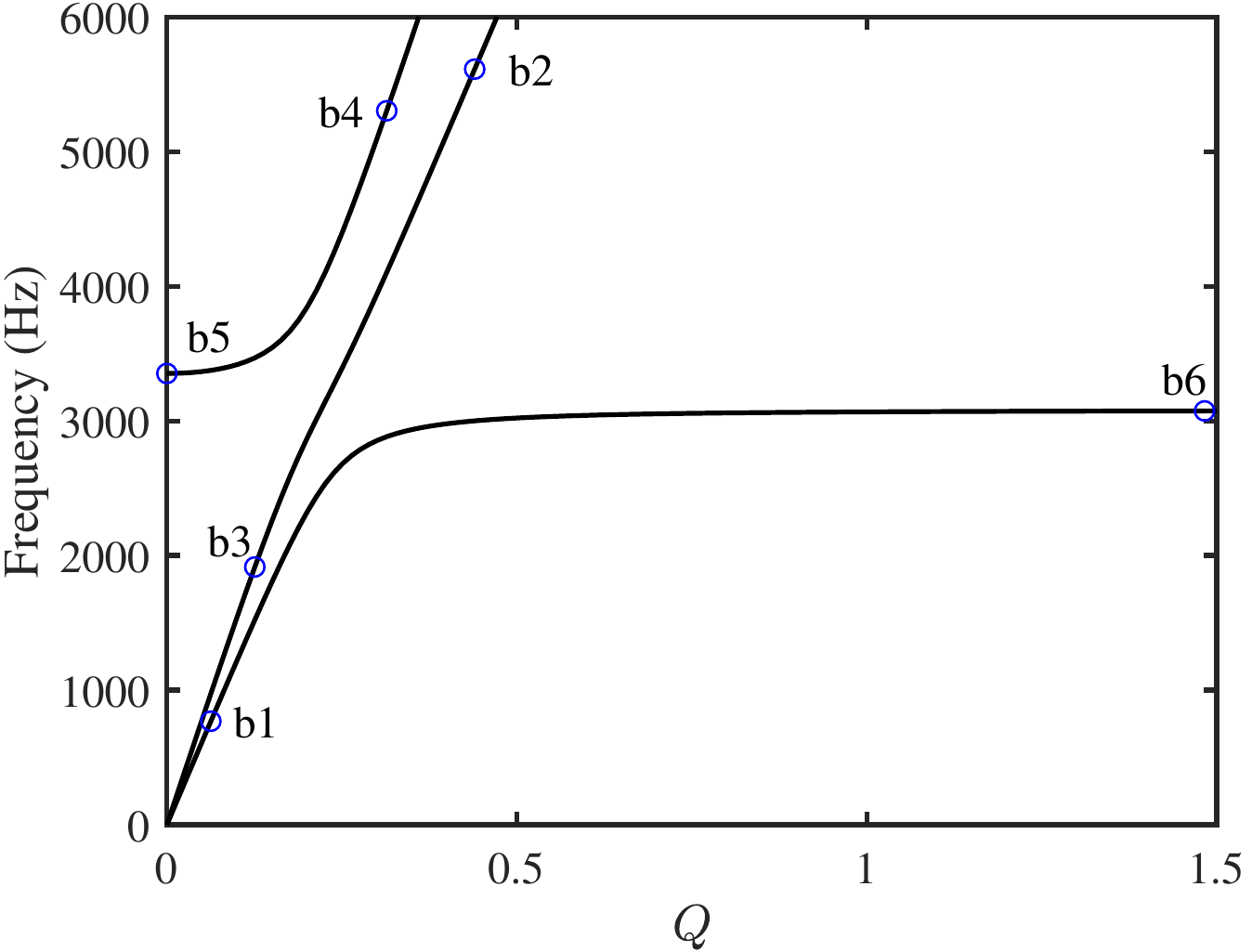}
\caption{$\theta = 45^\circ$ \label{fig:dc45comsoldots}}
\end{subfigure}\\
\begin{subfigure}[b]{0.5\linewidth}
\centering\includegraphics[height=175pt]{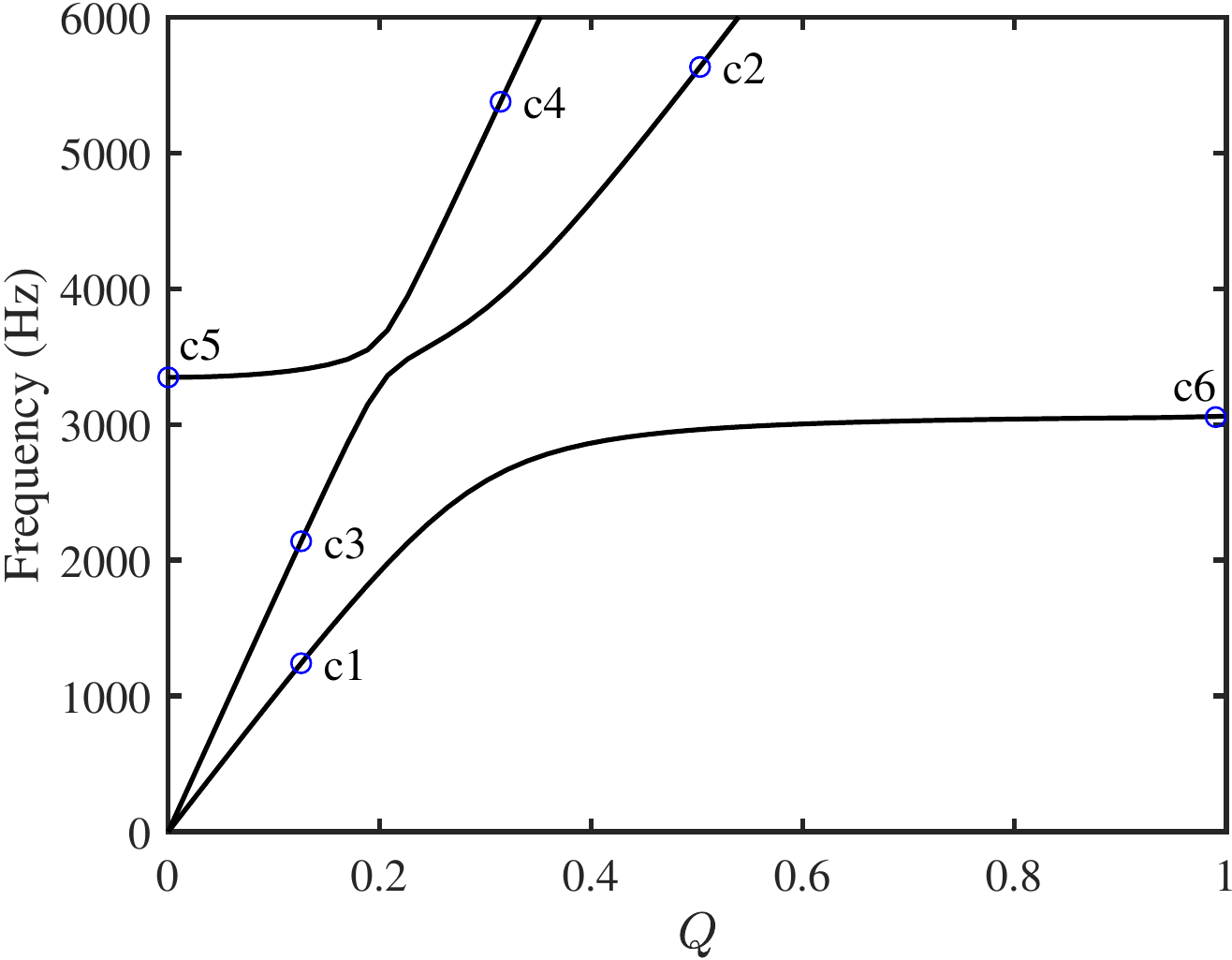}
\caption{$\theta = 60^\circ$ \label{fig:dc60comsoldots}}
\end{subfigure}%
\begin{subfigure}[b]{0.5\linewidth}
\centering\includegraphics[height=175pt]{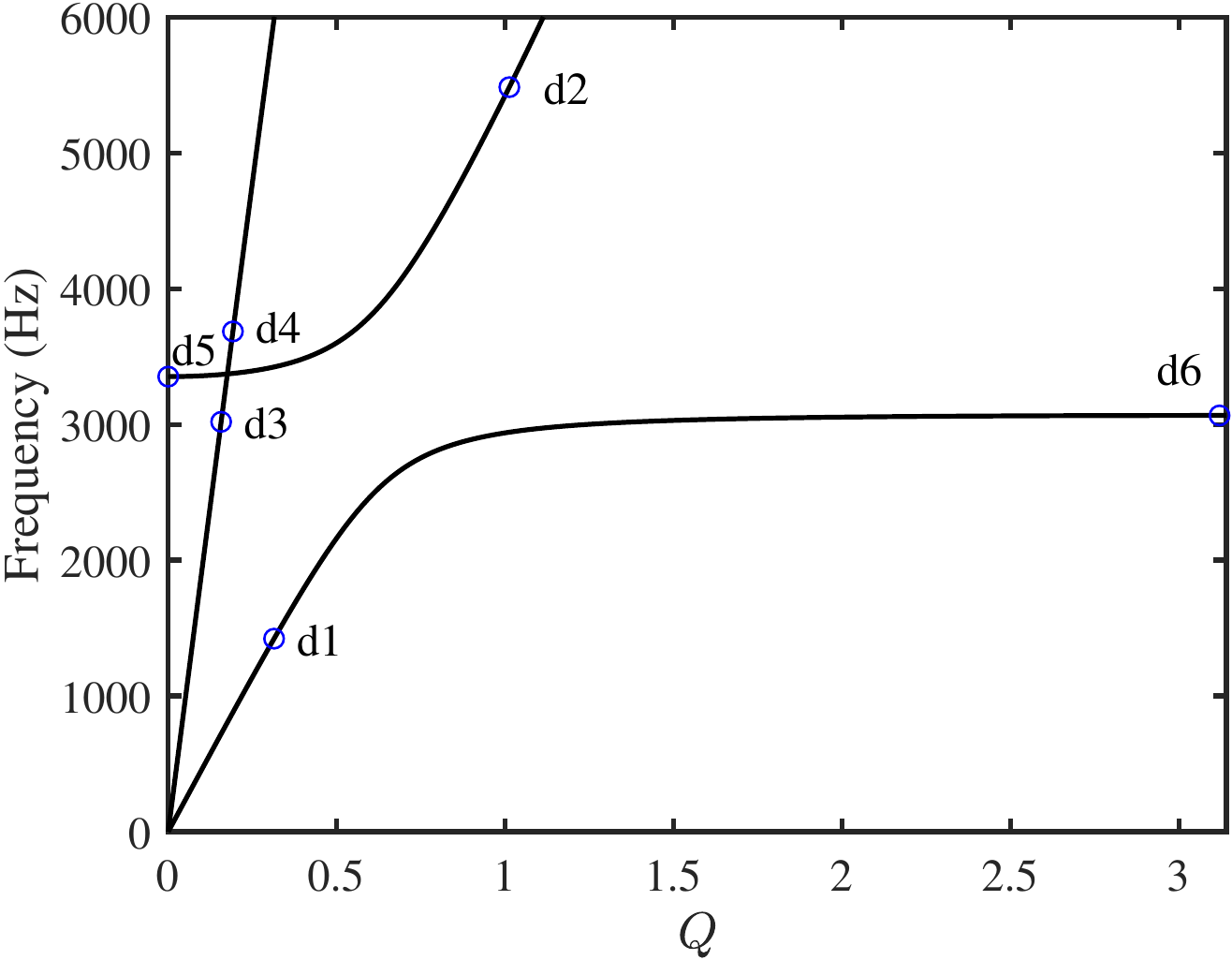}
\caption{$\theta = 90^\circ$ \label{fig:dc90comsoldots}}
\end{subfigure}%
\caption{\label{fig:dc}Dispersion curves for different propagation directions. Selected typical points are shown by small circles and labeled; see \cref{fig:modeshape}.} 
\end{figure}

\begin{figure}[!ht]
\begin{subfigure}[b]{0.5\linewidth}
\centering\includegraphics[height=150pt]{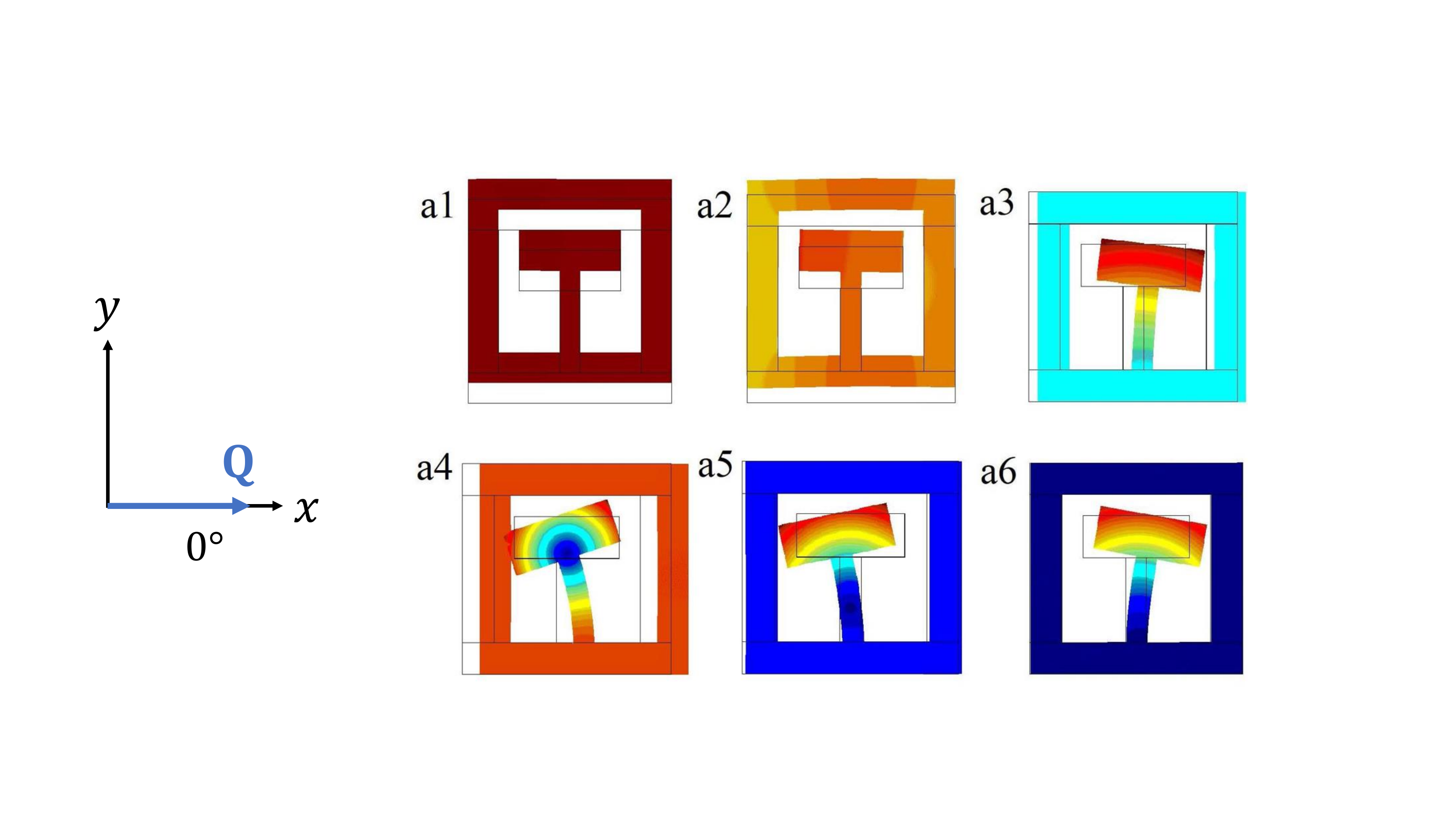}
\caption{\label{fig:amode}}
\end{subfigure}%
\begin{subfigure}[b]{0.5\linewidth}
\centering\includegraphics[height=150pt]{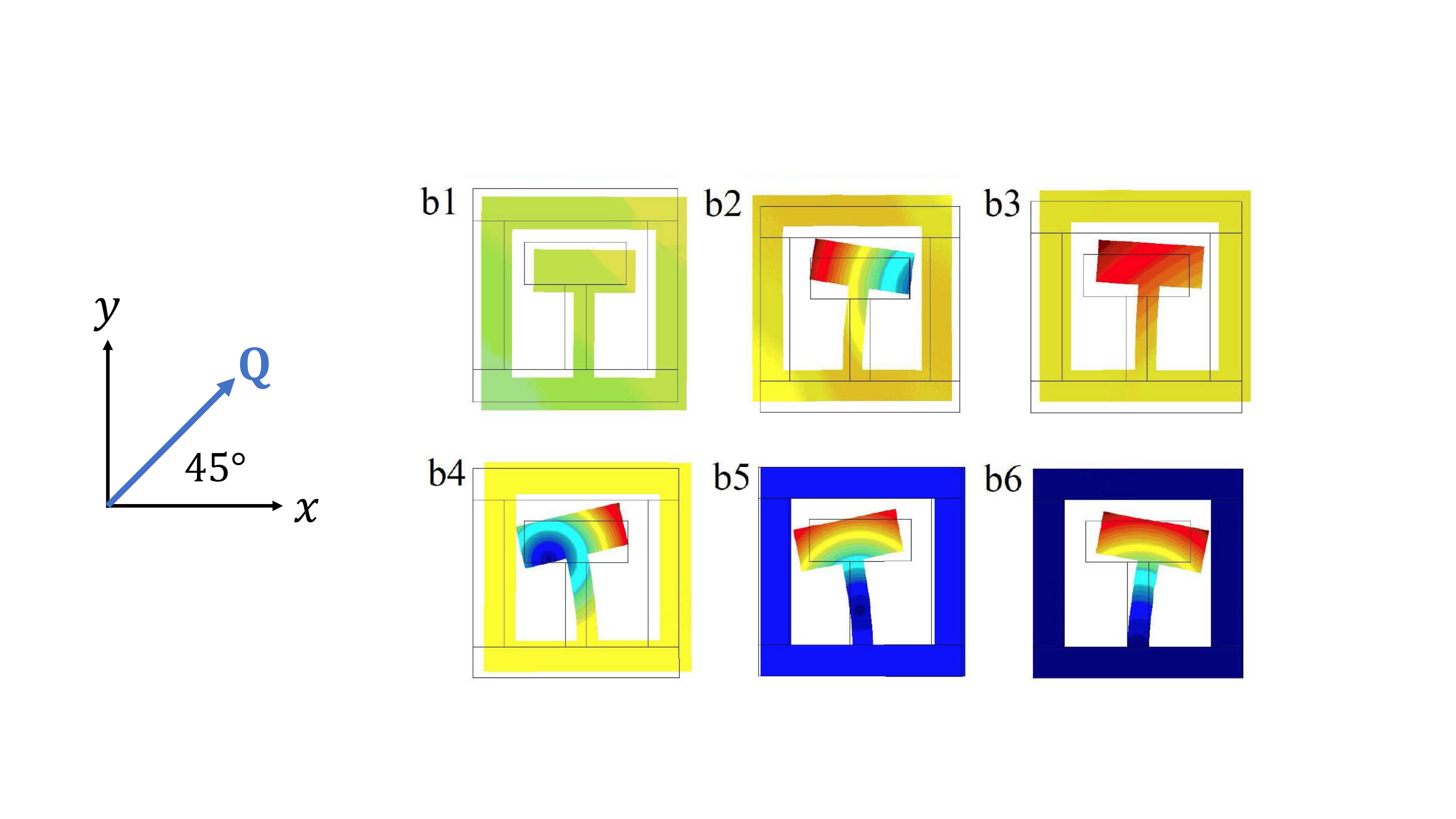}
\caption{\label{fig:bmode}}
\end{subfigure}\\
\begin{subfigure}[b]{0.5\linewidth}
\centering\includegraphics[height=150pt]{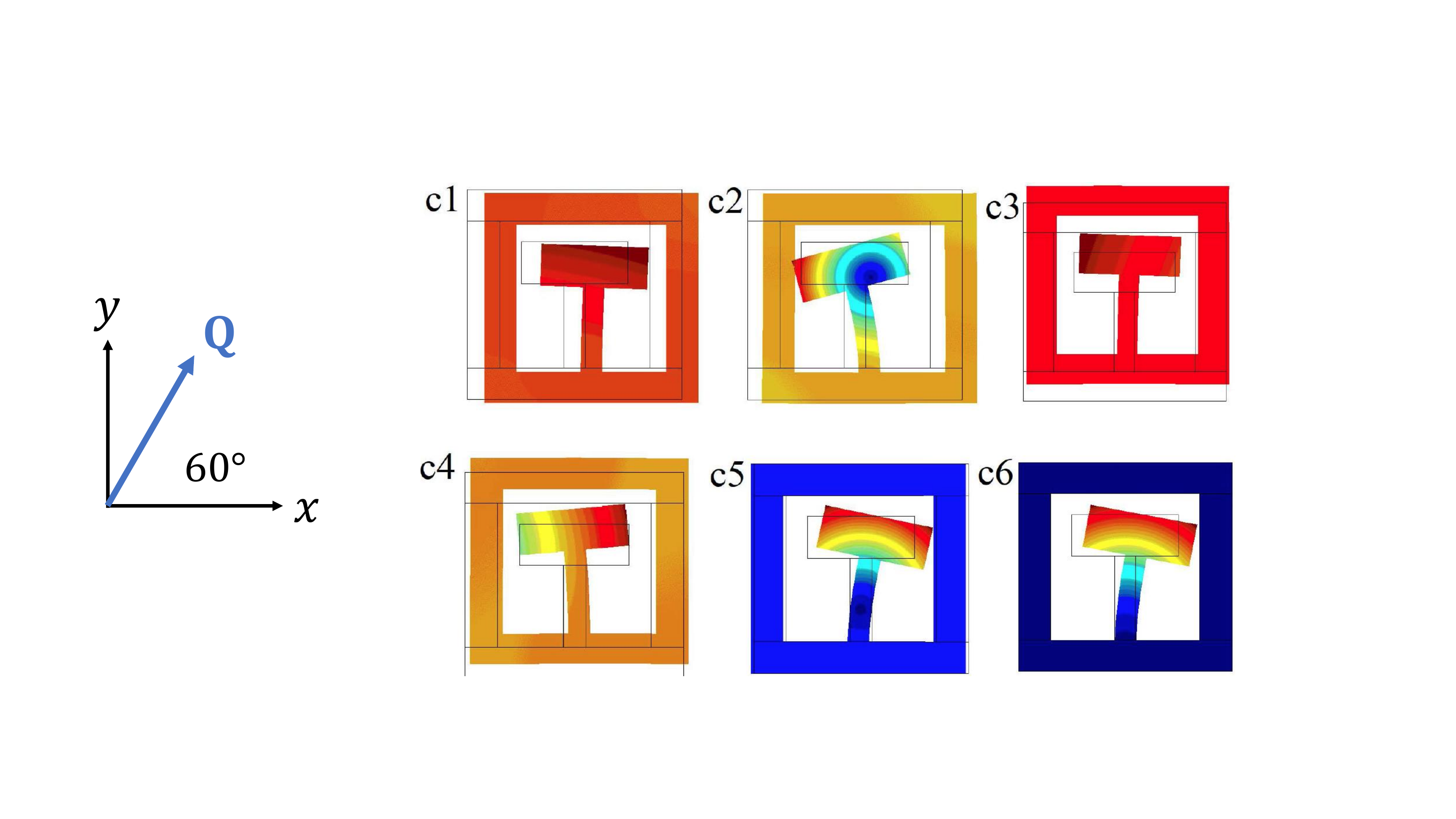}
\caption{\label{fig:cmode}}
\end{subfigure}%
\begin{subfigure}[b]{0.5\linewidth}
\centering\includegraphics[height=150pt]{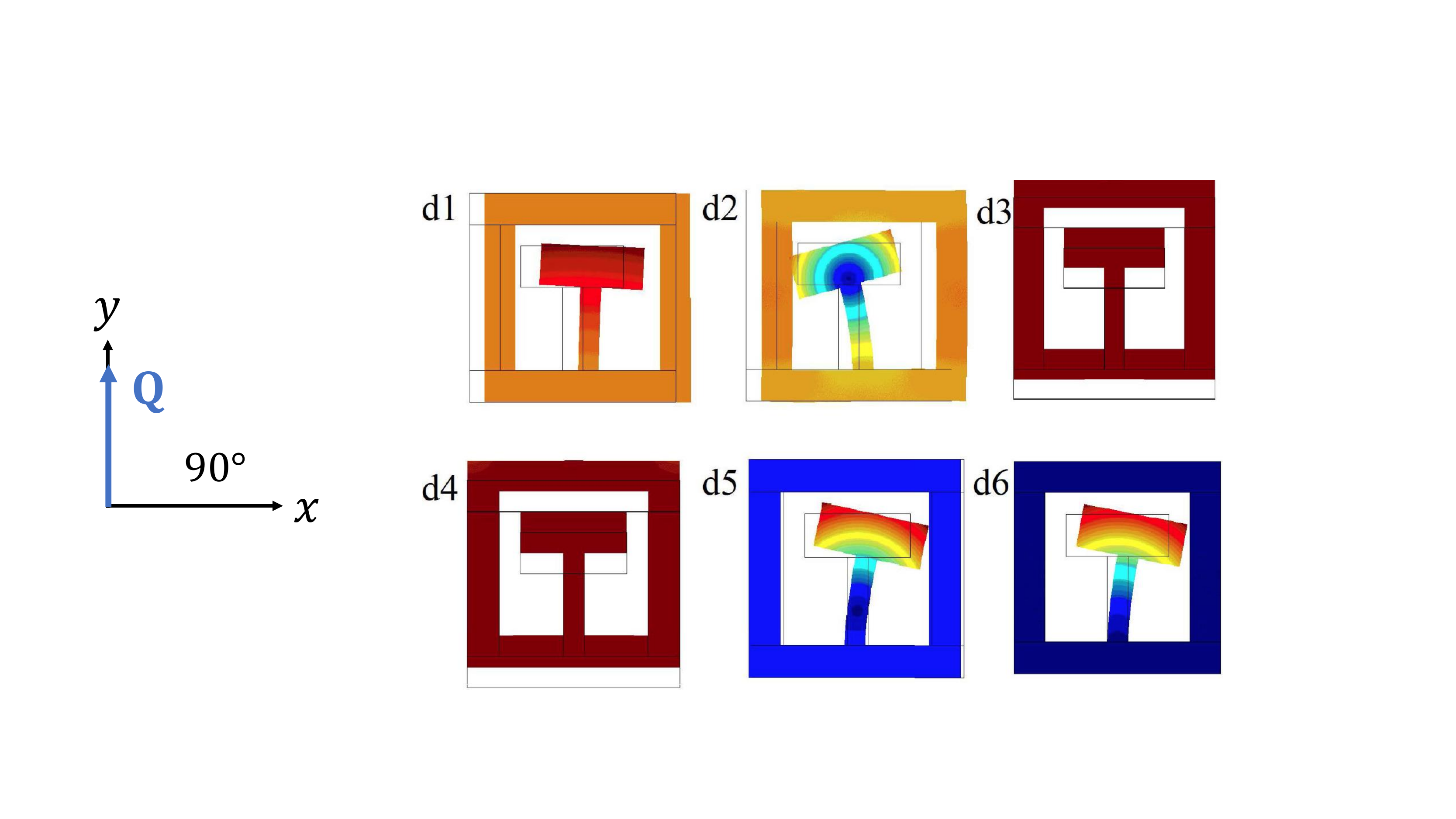}
\caption{\label{fig:dmode}}
\end{subfigure}
\caption{\label{fig:modeshape} Mode shapes of selected points in \cref{fig:dc}.} 
\end{figure}

The dispersion curves along specific directions can be obtained by fixing the propagation angle $\theta$ and plot the mode frequencies with respect to $Q$. \Cref{fig:dc} shows the dispersion curves for 0$^\circ$, 45$^\circ$, 60$^\circ$, and 90$^\circ$ directions. Typical points are identified by circles and labeled. The mode shapes of these points are shown in \cref{fig:modeshape}. Based on the observation of mode shapes, it can be concluded that the unit cell and its resonator have three main DOFs: the cell displacements in $x$ and $y$ directions and the lateral deflection of the resonator. Other DOFs can be observed to be independent in higher frequency ranges, which is beyond the considerations of the present work. The dispersion curves in \cref{fig:dc} and the mode shapes in \cref{fig:modeshape} indicate that there are mode mixing due to the coupling in the three main DOFs. Points a1, b1, c1, and d1 are examples of primarily acoustic shear vertical (SV) modes, while points a2, b2, c2, and d2 may be considered as optical SV modes. Points a3 and a4, b3 and b4, c3 and c4, d3 and d4 are examples of primarily longitudinal wave (P) modes. Points a5, b5, c5, and d5 represent a strong relative vibration between the resonator and cell, with opposing phases. Points a6, b6, c6, and d6 represent the local resonance modes where the T shaped cantilever absorbs most of the energy. Except for 90$^\circ$ direction shown in \cref{fig:dc90comsoldots}, the P wave acoustic branches are mixed together with SV wave optical branches. 

\section{Analytical Approach Based on Reduced Order Model}

\begin{figure}[!ht]
	\centering\includegraphics[width=350pt]{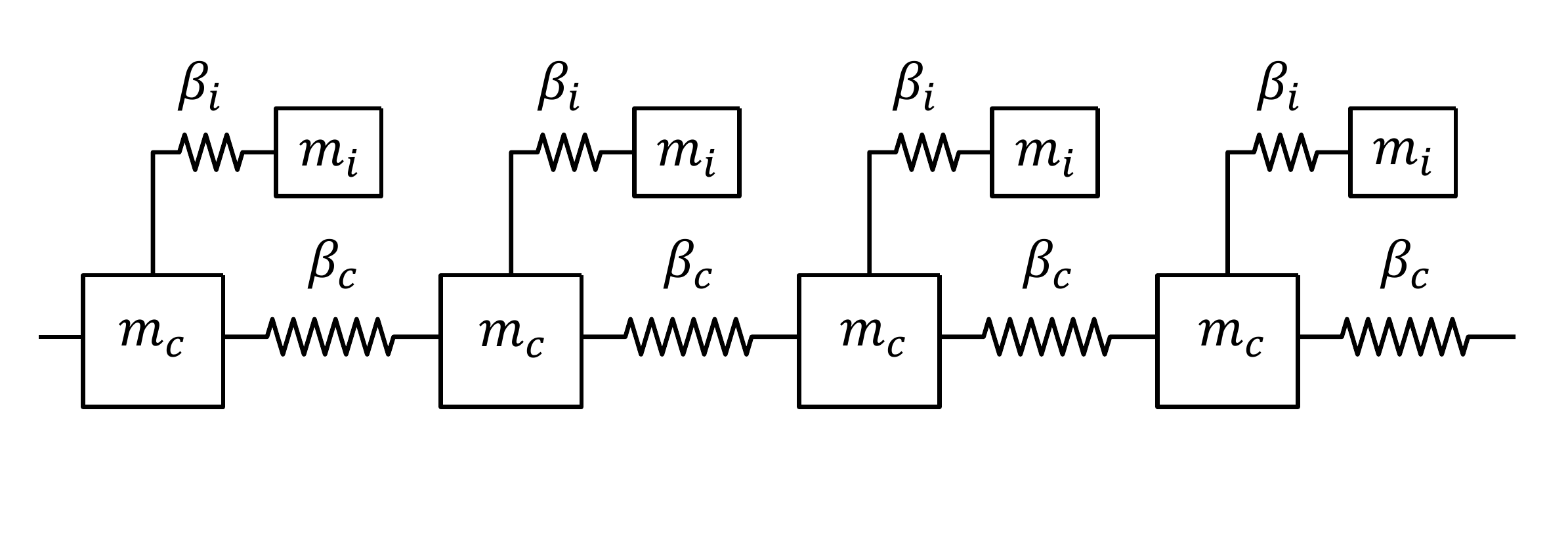}
	\caption{\label{fig:1dchain} Schematic of an infinite 1D discrete elastic chain with local resonators in each cell.}
\end{figure} 

\begin{figure}[!ht]
\begin{subfigure}[b]{0.5\linewidth}
\centering\includegraphics[height=180pt]{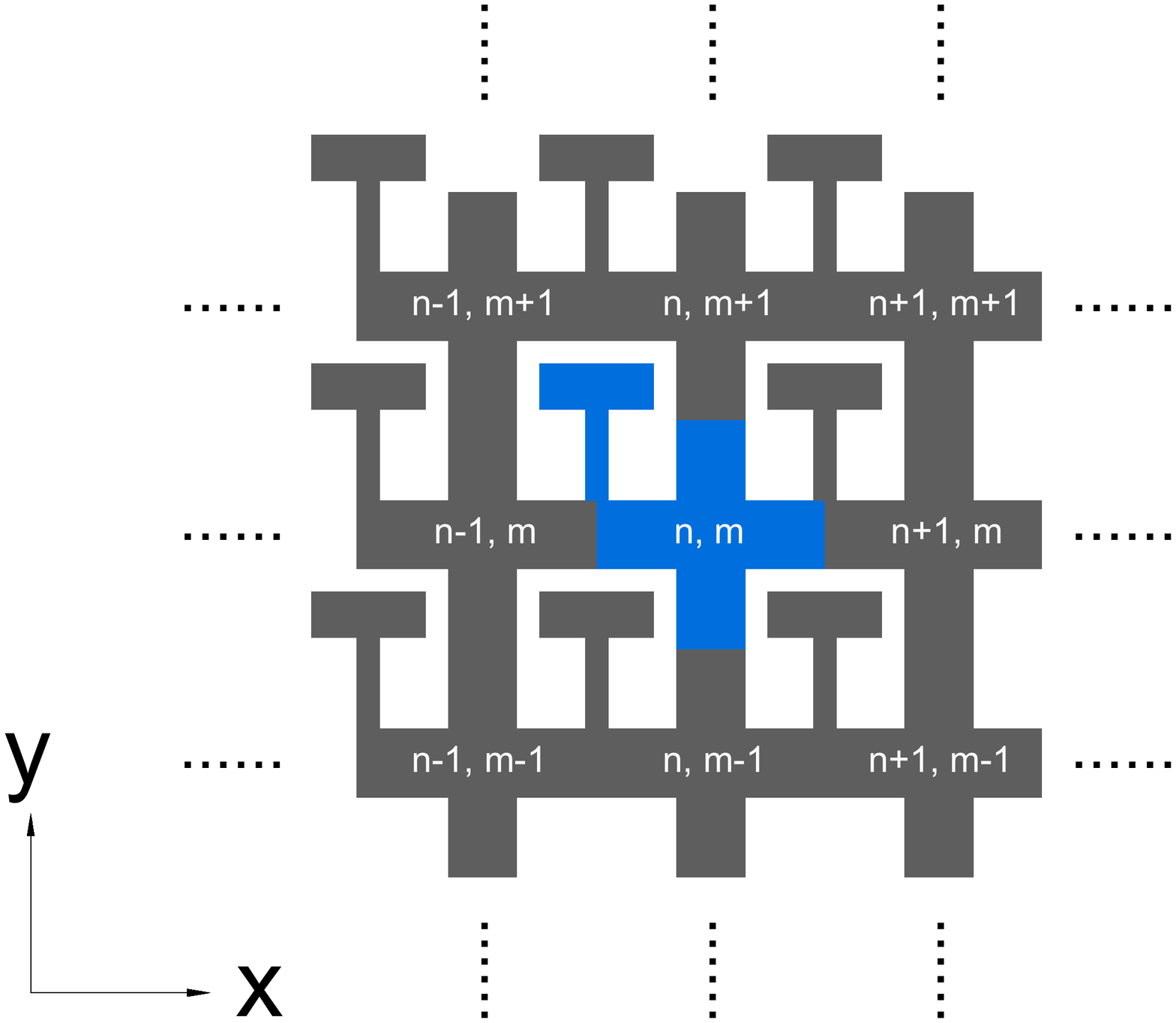}
\caption{\label{fig:schematic2}}
\end{subfigure}%
\begin{subfigure}[b]{0.5\linewidth}
\centering\includegraphics[height=160pt]{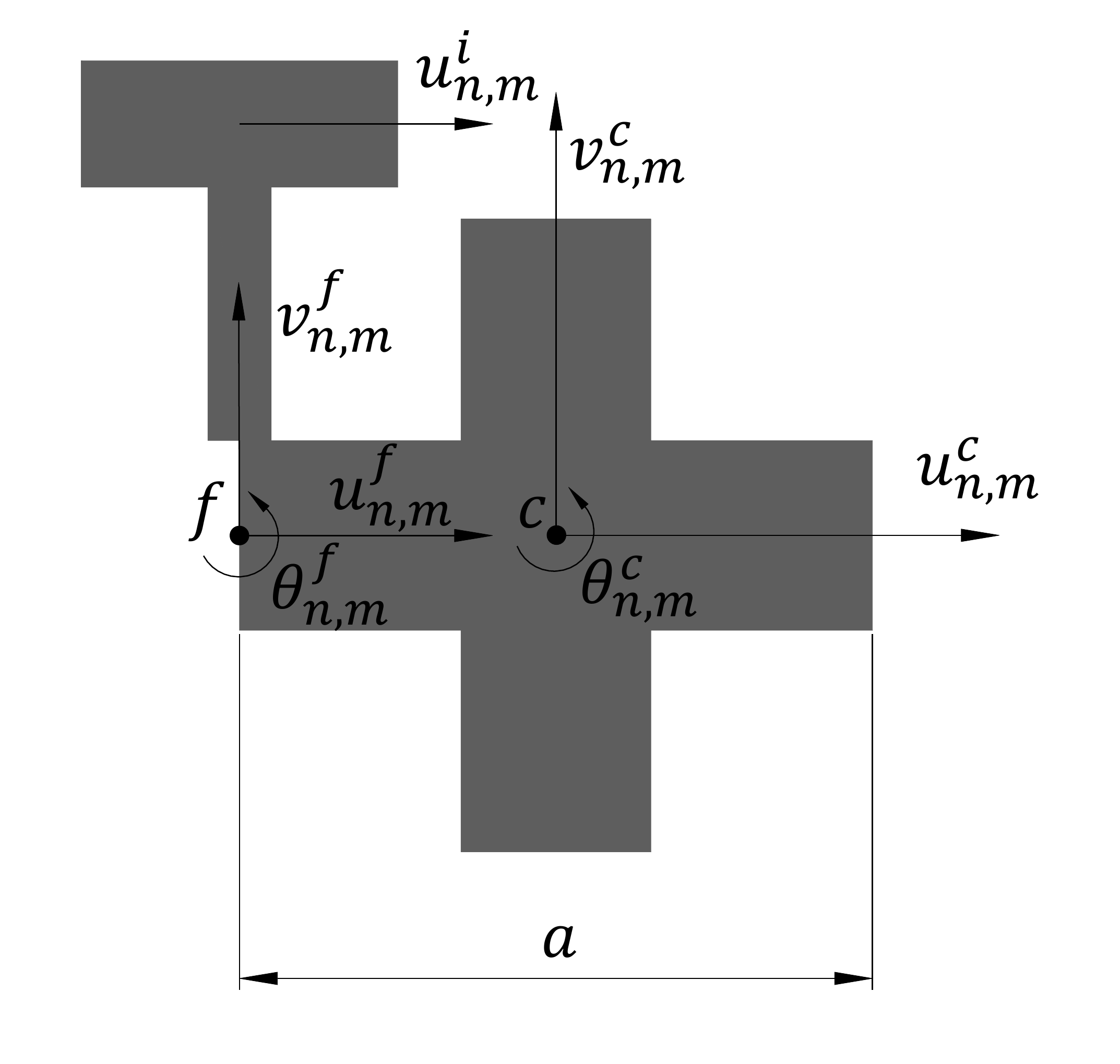}
\caption{\label{fig:ruc3}}
\end{subfigure}
\caption{\label{fig:structure2} (\subref{fig:schematic2}) Schematic drawing of the studied metamaterial with re-drawn RUC for reduced order analytical modeling. The $(n,\ m)^{\textrm{th}}$ cell is colored blue in the center. (\subref{fig:ruc3}) shows the geometry of the RUC with DOFs and reference points $c$ and $f$.}
\end{figure}

The general idea of the reduced order analytical approach is to discretize the repeating unit cell (RUC) into a finite number of structural elements and select degrees of freedoms based on the physical expectation of the resonator behavior. An oversimplified model for the T-resonator arrays in 1D is represented in \cref{fig:1dchain}. Here $m_c$ is associated with the mass of the frame in each cell. There is a spring with stiffness $\beta_c$ between two neighbor cells. A resonator with mass $m_i$ is connected to the each cell mass by a spring with stiffness $\beta_i$. The cell and resonator have horizontal DOFs $u^c_n$ and $u^i_n$, respectively. The integer $n$ represents the horizontal location of the cells and resonators in the chain and superscripts $c$ and $i$ represent cell/crystal and inclusion/resonator, respectively. The dynamic equations can be established for the $n^\mathrm{th}$ cell and resonator:
\begin{align}\label{eq:1dc}
m_c\frac{\partial^2u^c_n}{\partial t^2}&=\beta_c(u^c_{n+1}-2u^c_n+u^c_{n-1})+\beta_i(u^i_n-u^c_n), \\
\label{eq:1di}
m_i\frac{\partial^2u^i_n}{\partial t^2}&=\beta_i(u^c_n-u^i_n).
\end{align}
The band structure of the 1D chain model in normal incident P wave can be obtained by solving the characteristic equation based on \crefrange{eq:1dc}{eq:1di} and plotting the eigen-frequencies. The high number of variables in these equations is naturally reduced by using Bloch-Floquet periodicity. The conceptual approach may be expanded to 2D and 3D as well as more complex unit cell structures. However, in the case studied here a number of further assumptions are made based on the physical understanding of the system. The RUC is re-selected as a cross-shaped cell with a T-shaped cantilever attached, as shown in \cref{fig:structure2} (in contrast with numerical approach, where the a square frame and internal T-shaped cantilever were analyzed). With this selection, each cell is connected to neighbors through cuts transverse to the beam axes. The cross-shaped cell has two independent DOFs: displacement in $x$ direction is denoted by $u^c_{n,m}$, and $v^c_{n,m}$ represents its displacement in $y$ direction. The same could be done for the tip mass of the T-shaped cantilever, but as it turns out (based on the expected physics of the resonator) the vertical displacement may be eliminated geometrically and the only independent resonator DOF is the horizontal displacement $u^i_{n,m}$ at the center of tip mass in $x$ direction. Index $m$ represents the vertical location of the cells and resonators in the array.  

The trial displacement solutions that inherently satisfy Bloch-Floquet periodicity are:
\begin{equation}\label{eq:uc}u^c_{n,m}=u^c_0\ \mathrm{exp}\left[-i(\omega t-k_xna-k_yma)\right],
\end{equation}
\begin{equation}\label{eq:ui}u^i_{n,m}=u^i_0\ \mathrm{exp}\left[-i(\omega t-k_xna-k_yma)\right],
\end{equation}
\begin{equation}\label{eq:vc}v^c_{n,m}=v^c_0\ \mathrm{exp}\left[-i(\omega t-k_xna-k_yma)\right],
\end{equation}
where $u^c_0, u^i_0$ and $v^c_0$ are the complex displacement amplitudes to be determined along with $\omega$ for each pair values of $k_x$ and $k_y$. 
The relationship between forces and displacements of the RUC can be established by breaking it into elastic beams and rigid bodies. To obtain the force balance equations for the beams, one could introduce more extra dependent DOFs at reference points $c$ and $f$, as shown in \cref{fig:ruc3}.
Point $c$ is chosen at center of the cross. In addition to two primary DOFs $u^c$ and $v^c$, it also has a rotational DOF $\theta^c$ as the counter clockwise rotation angle. Point $f$ is defined at the intersection of the T-shaped resonator beam and cell frame's horizontal wall. 
Point $f$ has three dependent DOFs $u^f$, $v^f$, and $\theta^f$ defined in positive $x$, $y$, and counter clockwise directions, respectively. The four dependent DOFs defined at reference points will also need to satisfy Bloch-Floquet conditions. 

In the following calculations, all structural elements are treated as Timoshenko beams where shear deformations are allowed. In certain cases (where physically expected) the axial stiffness of the beams are also included in the process. The force balance equations of a Timoshenko beam element (in plane strain conditions) with four DOFs can be expressed as:
\begin{equation}\label{eq:tbm}
\frac{EI}{(1-\nu^2)(1+\Phi)L^3}
\left[
\begin{matrix}
12           &                                                             6L            &             -12          &             6L            \\
6L            &                             (4+\Phi)L^2        &             -6L          &             (2-\Phi)L^2         \\
-12          &                             -6L                                          &             12           &             -6L          \\
6L            &                             (2-\Phi)L^2         &             -6L          &             (4+\Phi)L^2
\end{matrix}
\right]
\left[
\begin{matrix} 
u_0 \\\phi_0\\u_1\\\phi_1
\end{matrix}
\right]=
\left[
\begin{matrix}
F_0\\M_0\\F_1\\M_1
\end{matrix}
\right],
\end{equation}
where the parameter $\Phi$ is defined as
\begin{equation} \Phi=\frac{12EI}{\kappa L^2GA(1-\nu^2)}~.\end{equation}
Here $\kappa=5/6$ is the shear correction factor for rectangular cross section, $E$ is Young's modulus of the material, $L$ is the length of beam, $A$ is the cross sectional area, $I$ is the second moment of inertia, and $G$ is the shear modulus. In \cref{eq:tbm}, the displacement $u_0$ and $u_1$ denote the transverse deflections at the two ends of the beam, and the angles $\phi_0$ and $\phi_1$ represent the rotation angles of cross section at these ends. The moments and angles at the two ends of beam are positive in counter clockwise direction.

\begin{figure}[!ht]
	\begin{subfigure}[c]{0.5\linewidth}
		\centering\includegraphics[height=160pt]{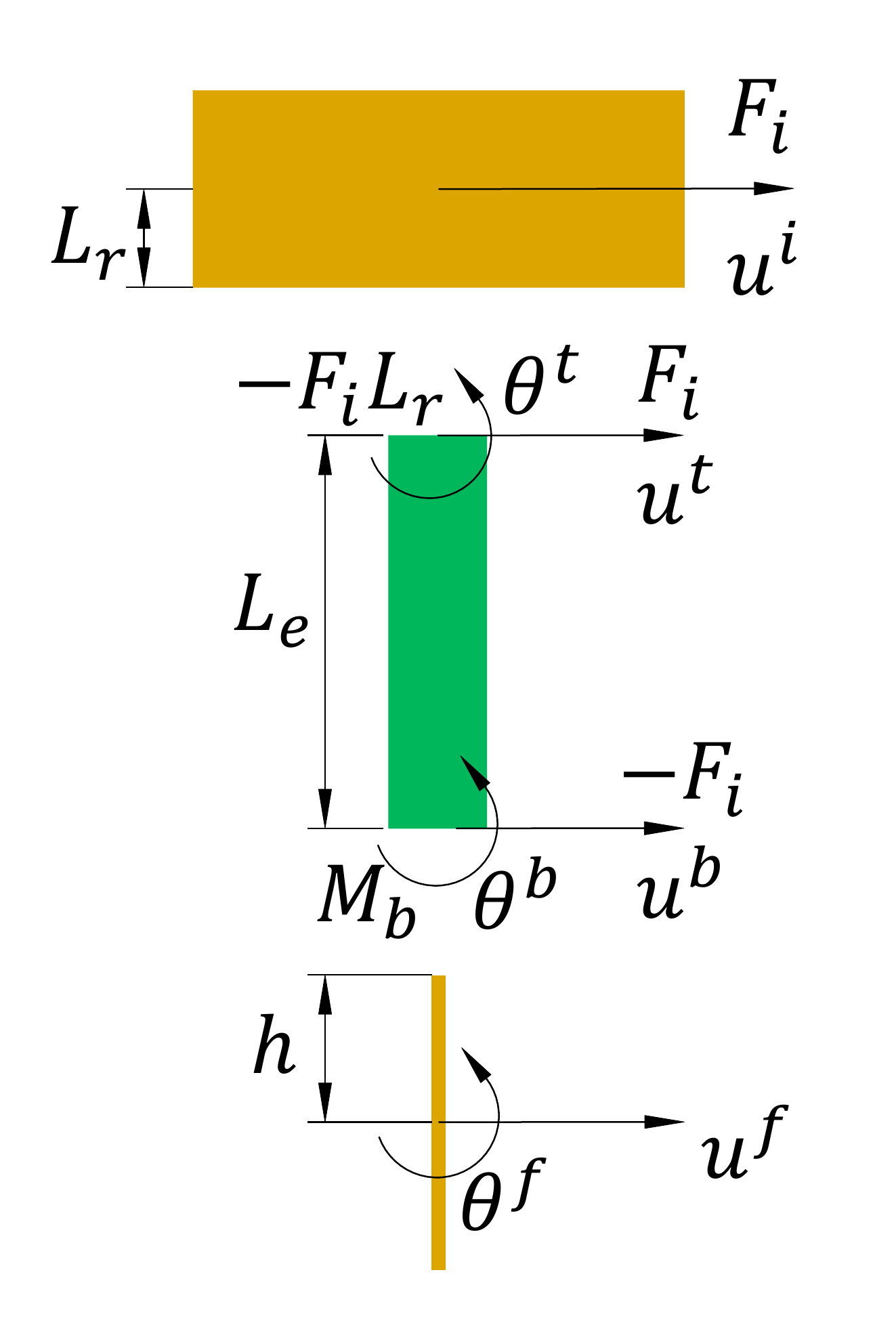}
		\caption{\label{fig:beta}}
	\end{subfigure}%
	\begin{subfigure}[c]{0.5\linewidth}
		\centering\includegraphics[height=160pt]{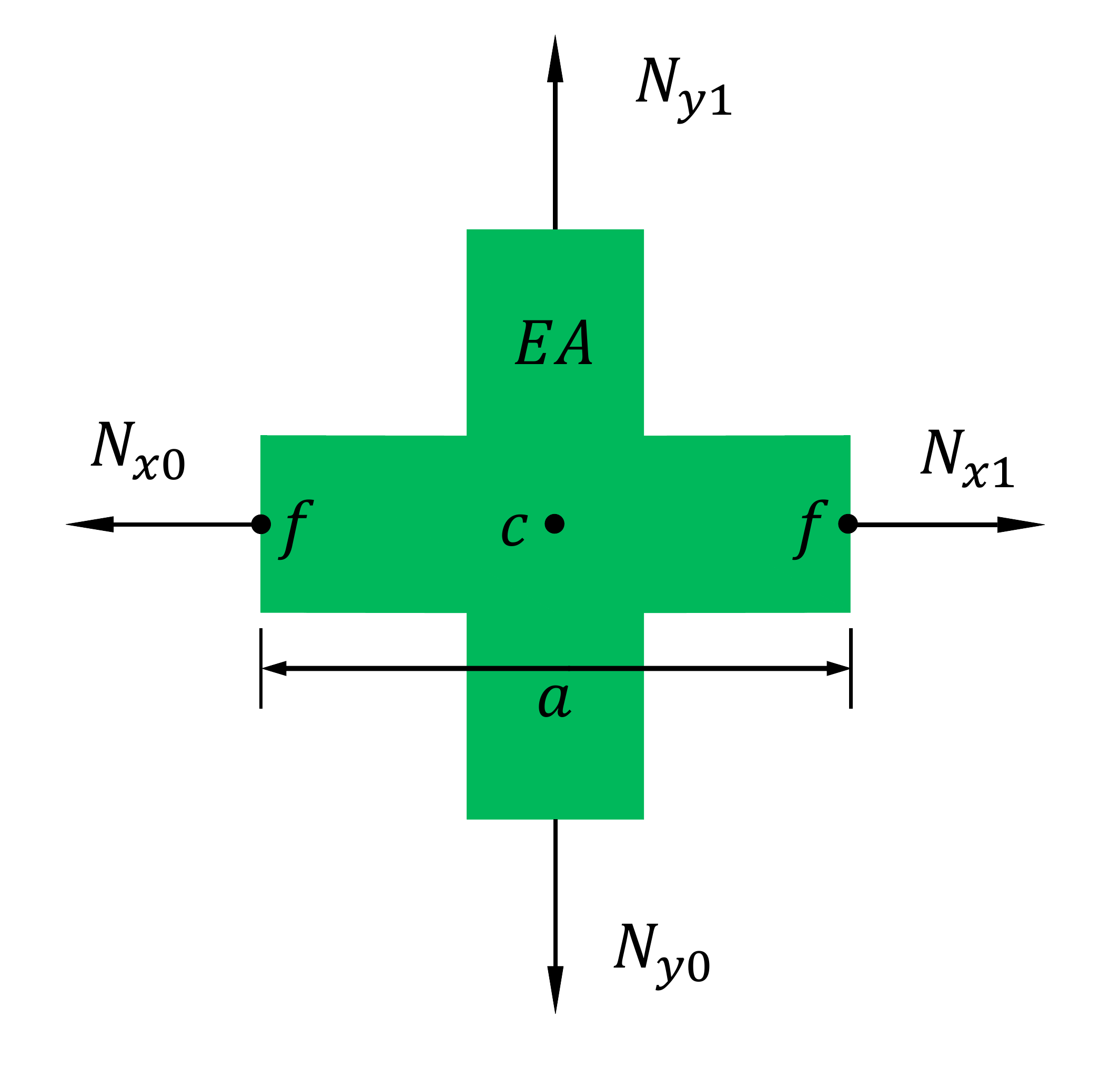}
		\caption{\label{fig:nforce}}
	\end{subfigure}%
	\caption{\label{fig:betaandnforce} (\subref{fig:beta}) Schematics of the resonator for effective stiffness calculation. All the parts are assumed rigid, except for the elastic Timoshenko vertical beam. Only $F_i$ and the loads on the main beam are shown. Note that the bottom part is assumed to have zero thickness and is presented to streamline the process of geometric elimination of extra DOFs. (\subref{fig:nforce}) Axial forces on the $(n,m)^\textrm{th}$ cell induced by relative displacements between points $c$ and $f$ as well as vertical neighbor cells. }
\end{figure}

\paragraph{Resonator effective mass and stiffness}
For mechanical metamaterials, the stop band location is mainly affected by the natural frequency of the resonator, which can be determined by resonator mass and its effective stiffness. The T-shaped resonator can be regarded as a rigid tip that stays on a vertical elastic beam, and its mass per unit thickness ($m_i = $ \SI{0.1161}{g/cm} in the example studied here) is chosen to be the resonator effective mass. To determine an equivalent spring-mass model, consider an imaginary concentrated force $F_i$ applied at the center of the rigid tip in positive $x$ direction as in \cref{fig:beta}. In order to find the effective resonator stiffness $\beta_i$, the balance equations for the vertical elastic beam can be written based on \cref{eq:tbm}:
\begin{equation}\label{eq:resobeam}
\frac{EI_e}{(1-\nu^2)(1+\Phi_e)L_e^3}
\left[
\begin{matrix}
12           &                      6L_e           &             -12          &             6L_e            \\
6L_e            &         (4+\Phi_e)L_e^2        &             -6L_e          &             (2-\Phi_e)L_e^2         \\
-12          &                             -6L_e                   &             12           &             -6L_e          \\
6L_e            &                    (2-\Phi_e)L_e^2         &             -6L_e          &             (4+\Phi_e)L_e^2
\end{matrix}
\right]
\left[
\begin{matrix} 
u^t \\\theta^t\\u^b\\\theta^b
\end{matrix}
\right]=
\left[
\begin{matrix}
F_i\\-F_iL_r\\-F_i\\M_b
\end{matrix}
\right],
\end{equation}
where subscript $e$ refers to the properties of the elastic beam under the rigid tip, $u^t$ and $u^b$ represent the displacements at top and bottom ends of the beam in positive $x$ direction, $\theta^t$ and $\theta^b$ are the rotation angles, $L_r$ ($= $ \SI{1}{mm}) is the distance from center of tip mass to top of the vertical beam, and $M_b$ is the resultant moment at the bottom. Note the value of moment at top of the beam is known and given based on the moment equilibrium of the tip mass and the assumption that its moment of inertia is negligible in this problem. As will be seen in the following, the unbalanced moment induced in the beam is in fact an important and fundamental component of the unusual behavior of the system. However, the simplifying assumption of the local equilibrium for the tip appears not to limit this effect in any significant way. The kinematic relations among the DOFs are 
\begin{equation}\label{eq:uiut}u^i=u^t-\theta^t L_r~,\end{equation}
\begin{equation}\label{eq:ubuf}u^b=u^f-\theta^f h,\end{equation}
\begin{equation}\label{eq:tbtf}\theta^b=\theta^f~,\end{equation}
where $h$ ($= $ \SI{1.5}{mm}) is half of the cell wall thickness. Substituting \crefrange{eq:uiut}{eq:tbtf} into \cref{eq:resobeam} yields:
\begin{equation}
F_i=\beta_i\left[u^i-u^f+\theta^f(L_e+L_r+h)\right],
\end{equation}
\begin{equation}M_b=F_i\left(L_e+L_r\right),\end{equation}
where the effective resonator stiffness is:
\begin{equation}\beta_i=\frac{12EI_e}{((4+\Phi_e)L_e^3+12L_e^2L_r+12L_eL_r^2)(1-\nu^2)}.
\end{equation}
For the cell geometry studied here (assuming \SI{1}{cm} unit thickness) $\beta_i \approx $ \SI{66945.4}{(N/cm)/m}. 


\paragraph{Frame dynamics}
For the cross-shaped cell, it's assumed that its mass ($m_c = $ \SI{0.63855}{g/cm} for the example studied here) is concentrated at the center point $c$.
To study the horizontal dynamics of the cell, we assume that its structure is fully elastic and the cross-sectional area $A$ is uniform along the length between reference points by neglecting the lateral constraints at the ends of beams. 
The axial forces applied to the $(n,m)^\textrm{th}$ cell , see \cref{fig:nforce}, can be expressed as
\begin{equation}
N_{y1}=\frac{EA}{(1-\nu^2)a}\left(v^c_{n,m+1}-v^c_{n,m}\right),
\end{equation}
\begin{equation}
N_{y0}=\frac{EA}{(1-\nu^2)a}\left(v^c_{n,m}-v^c_{n,m-1}\right),
\end{equation}
\begin{equation}
N_{x1}=\frac{2EA}{(1-\nu^2)a}\left(u^f_{n+1,m}-u^c_{n,m}\right),
\end{equation}
\begin{equation}
N_{x0}=\frac{2EA}{(1-\nu^2)a}\left(-u^f_{n,m}+u^c_{n,m}\right).
\end{equation}
Note that the vertical DOF on the top and bottom of the cross shaped cell has been eliminated based on the neighboring cell DOFs; See \cref{fig:nforce}. The DOF at $f$, however, may not be eliminated like this due to the influence of the T-shaped resonator. 

\begin{figure}[!ht]
	\begin{subfigure}[b]{0.5\linewidth}
		\centering\includegraphics[height=106pt]{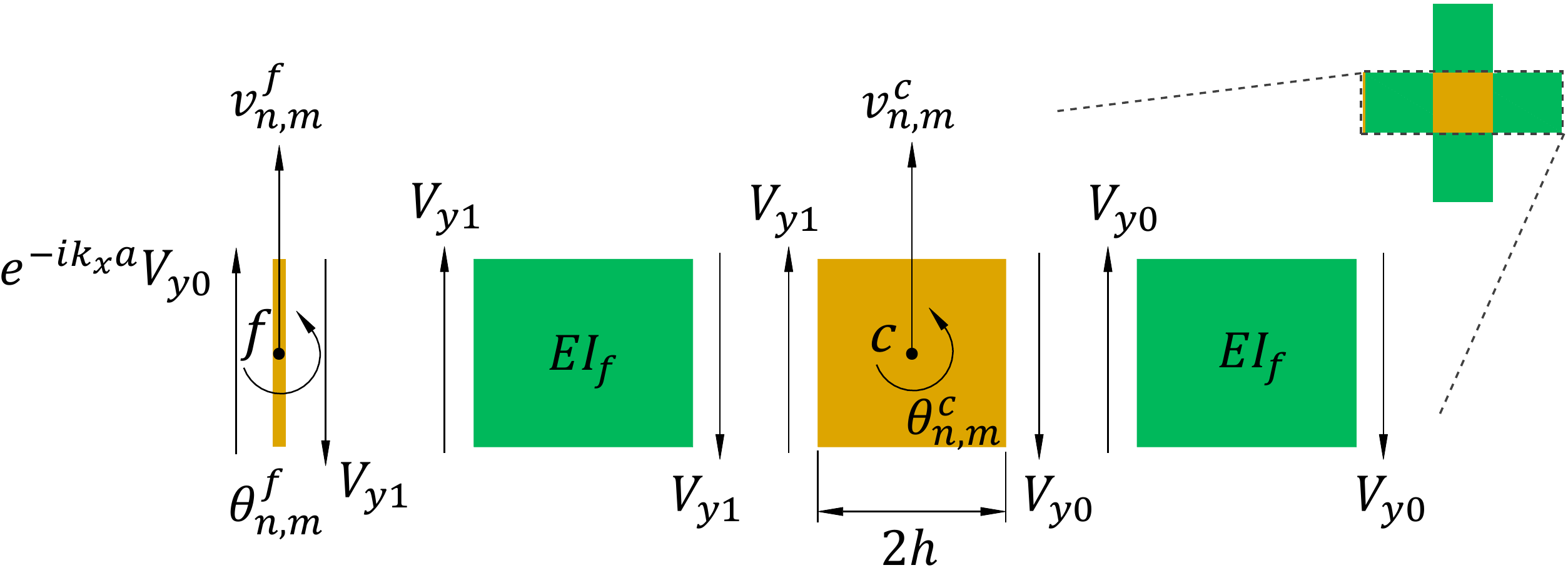}
		\caption{\label{fig:vys}}
	\end{subfigure}%
	\begin{subfigure}[b]{0.5\linewidth}
		\centering\includegraphics[height=106pt]{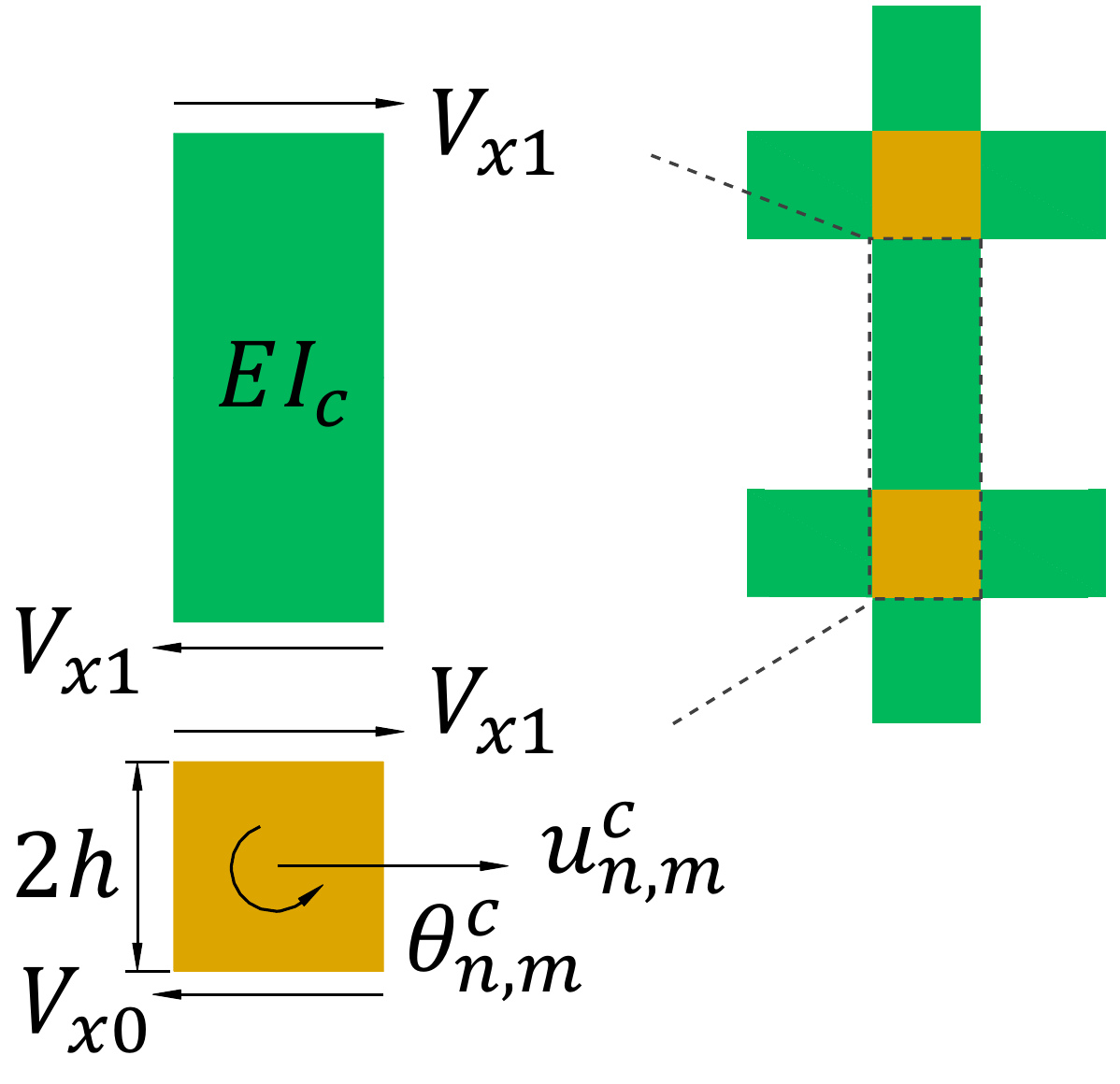}
		\caption{\label{fig:vxs}}
	\end{subfigure}
	\caption{\label{fig:vfs} The shear forces of $(n,m)^\textrm{th}$ cell in $y$ and $x$ directions, associated with (\subref{fig:vys})~horizontal and (\subref{fig:vxs}) vertical cell walls, respectively. The parts that include points $c$ and $f$ in this model are assumed to be rigid. The narrow part with point $f$ in (\subref{fig:vys}) is the same as the one shown in \cref{fig:beta}.}
\end{figure}

There are also shear forces applied to the cell due to beam deformations, as shown in \cref{fig:vfs}. Note the selection of rigid portions in this model. Subscript $c$ denotes the vertical beam (cell wall) and has length $L_c = a - 2h$. Subscript $f$ denotes the horizontal beam between the narrow part including node $f$ and the square at te center of the cross, containing node $c$, with length $L_f = a/2 - h$. The shear forces applied to the $(n,m)^\textrm{th}$ cell can be written based on Timoshenko beam theory: 
\begin{equation}
\label{eq:vy1}
V_{y1} = \vec{B}^V_f
\left[
\begin{matrix} 
v^f_{n,m}\\
\theta^f_{n,m}\\
v^c_{n,m}-\theta^c_{n,m}h\\
\theta^c_{n,m}
\end{matrix}
\right],
\end{equation}
\begin{equation}
\label{eq:vy0}
V_{y0} = -\vec{B}^V_f 
\left[
\begin{matrix} 
v^f_{n+1,m}\\
-\theta^f_{n+1,m}\\
v^c_{n,m}+\theta^c_{n,m}h\\
-\theta^c_{n,m}
\end{matrix}
\right],
\end{equation}
\begin{equation}
\label{eq:vx1}
V_{x1} = \vec{B}^V_c
\left[
\begin{matrix} 
u^c_{n,m+1}+\theta^c_{n,m+1}h\\
\theta^c_{n,m+1}\\
u^c_{n,m}-\theta^c_{n,m}h\\
\theta^c_{n,m}
\end{matrix}
\right],
\end{equation}
\begin{equation}
\label{eq:vx0}
V_{x0} = -\vec{B}^V_c
\left[
\begin{matrix} 
u^c_{n,m-1}-\theta^c_{n,m-1}h\\
-\theta^c_{n,m-1}\\
u^c_{n,m}+\theta^c_{n,m}h\\
-\theta^c_{n,m}
\end{matrix}
\right],
\end{equation}
where
\begin{equation}
	\vec{B}^V_\alpha = \frac{EI_\alpha}{(1+\Phi_\alpha)(1-\nu^2)L^3_{\alpha}}\left[
	\begin{matrix}
	12 & 6L_{\alpha} & -12 & 6L_{\alpha}
	\end{matrix}
	\right],
\end{equation}
with $\alpha = f, c$. 
Note that in \crefrange{eq:vy1}{eq:vx0}, the displacements at the beam ends are expressed not only in terms of $u^c$ and $v^c$, but also $\theta^c$ and $h$. This is due to kinematic relations based on rigidity of the relevant parts. Such assumption will inevitably increase the calculated SV wave velocity (the slope of SV wave branch). Another simpler assumption is that every point in the rigid portion have the same DOFs, and consequently in \crefrange{eq:vy1}{eq:vx0} the terms with $h$ in the displacement vectors will no longer appear. This will lead to a further overestimation of the SV wave speed. On the other hand, if we assume the effective beam lengths reach all the way through previously assumed rigid parts, e.g. to point $c$ at the center of square, the formulations may be rewritten with $h = 0$ and the SV wave speed will become much lower than numerical simulation results. The presented model leads to relatively accurate results while keeping the process as simple as possible. 

\begin{figure}[!ht]
	\begin{subfigure}[c]{0.5\linewidth}
		\centering\includegraphics[height=120pt]{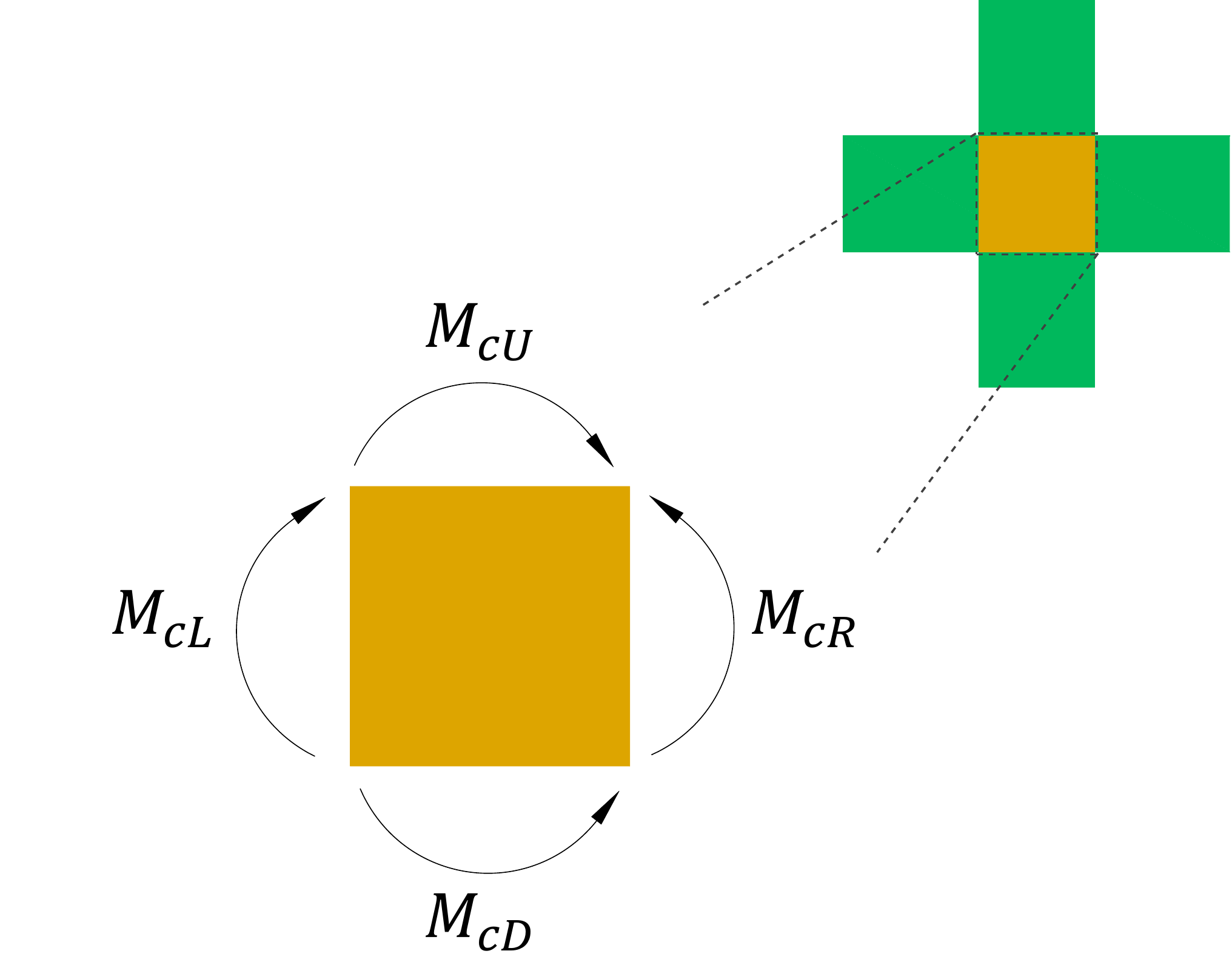}
		\caption{\label{fig:mc}}
	\end{subfigure}%
	\begin{subfigure}[c]{0.5\linewidth}
		\centering\includegraphics[height=130pt]{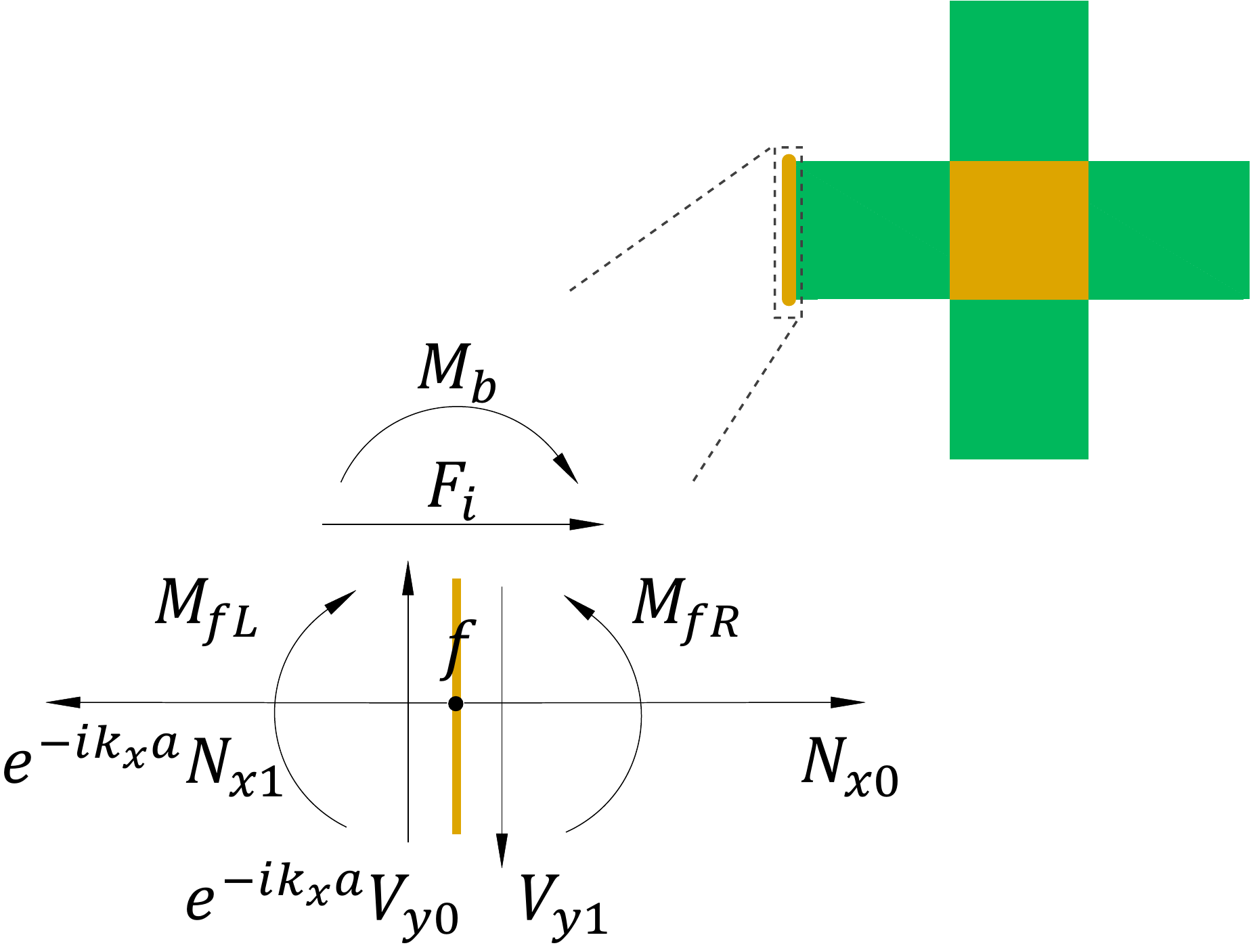}
		\caption{\label{fig:fbalance}}
	\end{subfigure}\\
	\caption{\label{fig:mcandfb} The moments applied on the square edges in $(n,m)^\textrm{th}$ cell are shown in (\subref{fig:mc}) with directions labeled. (\subref{fig:fbalance}) shows the free body diagram for the vertical rigid line where point $f$ stays in the middle. The yellow line shown is at the left side of $(n,m)^\textrm{th}$ cross-shaped cell.}
\end{figure}

The moments applied at the left, right, top, and bottom sides of the rigid square containing point $c$ in $(n,m)^\textrm{th}$ cell are denoted as $M_{cL}$, $M_{cR}$, $M_{cU}$, and $M_{cD}$, respectively, as shown in \cref{fig:mc}. 
Using the similar process as the one used for shear forces, the moments applied to the square can be written as
\begin{equation}
M_{cL} = \vec{B}^M_f
\left[
\begin{matrix} 
v^f_{n,m}\\
\theta^f_{n,m}\\
v^c_{n,m}-\theta^c_{n,m}h\\
\theta^c_{n,m}
\end{matrix}
\right],
\end{equation}
\begin{equation}
M_{cR} = - \vec{B}^M_f
\left[
\begin{matrix} 
-v^f_{n+1,m}\\
\theta^f_{n+1,m}\\
-v^c_{n,m}-\theta^c_{n,m}h\\
\theta^c_{n,m}
\end{matrix}
\right],
\end{equation}
\begin{equation}
M_{cU} = \vec{B}^M_c
\left[
\begin{matrix} 
u^c_{n,m+1}+\theta^c_{n,m+1}h\\
\theta^c_{n,m+1}\\
u^c_{n,m}-\theta^c_{n,m}h\\
\theta^c_{n,m}
\end{matrix}
\right],
\end{equation}
\begin{equation}
M_{cD} = - \vec{B}^M_c
\left[
\begin{matrix} 
-u^c_{n,m-1}+\theta^c_{n,m-1}h\\
\theta^c_{n,m-1}\\
-u^c_{n,m}-\theta^c_{n,m}h\\
\theta^c_{n,m}
\end{matrix}
\right],
\end{equation}
where
\begin{equation}
	\vec{B}^M_\alpha = \frac{EI_\alpha}{(1+\Phi_\alpha)(1-\nu^2)L^3_{\alpha}}\left[
	\begin{matrix}
	6L_{\alpha} & (2-\Phi_\alpha)L^2_{\alpha} & -6L_{\alpha} & (4+\Phi_\alpha)L^2_{\alpha}
	\end{matrix}
	\right], 
\end{equation}
with $\alpha = f, c$. 
The moments applied at the left and right sides of reference point $f$ are denoted as $M_{fL}$ and $M_{fR}$, as shown in \cref{fig:fbalance}, and they can be written as:


\begin{equation}
M_{fL} = \vec{B}^M_f
\left[
\begin{matrix} 
v^c_{n-1,m}+\theta^c_{n-1,m}h\\
\theta^c_{n-1,m}\\
v^f_{n,m}\\
\theta^f_{n,m}
\end{matrix}
\right],
\end{equation}
\begin{equation}
M_{fR} = -\vec{B}^M_f
\left[
\begin{matrix} 
-v^c_{n,m}+\theta^c_{n,m}h\\
\theta^c_{n,m}\\
-v^f_{n,m}\\
\theta^f_{n,m}
\end{matrix}
\right].
\end{equation}

All previous calculations are based on elements in the $(n,\ m)^{th}$ cell, and all the loads required for deriving dynamic equations of the RUC have been expressed. Note that the yellow line shown in \cref{fig:fbalance} is massless and has no width, therefore the forces and moments applied to it should be balanced statically. We also assume that the moments around point $c$ are balanced, i.e. ignore its moment of inertia. Combined with Bloch-Floquet periodicity:
\begin{equation} 
V_{y1}=e^{-ik_xa}V_{y0}~,
\end{equation}
the dependent DOF $v^f$ can be eliminated and expressed as
\begin{equation} 
v^f_{0}= \frac{e^{-ik_xa}+1}{2}v^c_0+\frac{e^{-ik_xa}-1}{4}(2h+L_f)\theta^c_0~.
\end{equation}
The equations of motion for the three primary DOFs $u^c$, $u^i$, and $v^c$ can now be written as
\begin{align}
\label{eq:ucm}m_c\frac{\partial^2 u^c_{n,m}}{\partial t^2}&=N_{x1}-N_{x0}+V_{x1}-V_{x0}~,\\
\label{eq:uim}m_i\frac{\partial^2 u^i_{n,m}}{\partial t^2}&=-F_i~,\\
\label{eq:vcm}\left(m_c+m_i\right)\frac{\partial^2 v^c_{n,m}}{\partial t^2}&=N_{y1}-N_{y0}+V_{y1}-V_{y0}~,
\end{align}
The remaining DOFs, $u^f$, $\theta^f$, and $\theta^c$ , may be eliminated using other static balance equations at the reference points: 
\begin{equation}
\label{eq:ufs}
F_i+N_{x0}-e^{-ik_xa}N_{x1}=0,
\end{equation}
\begin{equation}
\label{eq:thetafs}
M_{fL}-M_{fR}+M_b+F_ih=0,
\end{equation}
\begin{equation}
\label{eq:thetacs}
M_{cL}+M_{cU}-M_{cR}-M_{cD}+h(V_{x0}+V_{x1}+V_{y0}+V_{y1})=0.
\end{equation}
Alternatively and in general, the displacement vectors for the three primary DOFs and three dependent DOFs can be denoted as $\vec{U_p}$ and $\vec{U_d}$, respectively:
\begin{align}
\vec{U_p}&=\left[\begin{matrix}u^c_0&u^i_0&v^c_0\end{matrix}\right]^\top, \\
\vec{U_d}&=\left[\begin{matrix}u^f_0&\theta^f_0&\theta^c_0\end{matrix}\right]^\top.
\end{align}
Combining the three equations of motion \crefrange{eq:ucm}{eq:vcm} and the three balance equations \crefrange{eq:ufs}{eq:thetacs}, we obtain:
\begin{equation}
\label{eq:6by6}
\left[\begin{matrix}
\vec{D_{pp}} & \vec{K_{pd}}\\
\vec{K_{dp}}&\vec{K_{dd}}
\end{matrix}\right]
\left[\begin{matrix}
\vec{U_p}\\\vec{U_d}\end{matrix}\right]=0,
\end{equation}
where $\vec{K_{pd}}$, $\vec{K_{dp}}$, and $\vec{K_{dd}}$ are $3\times3$ stiffness sub-matrices, and $\vec{D_{pp}}$ is the $3\times3$ diagonal dynamic stiffness sub-matrix, which includes $-\omega^2 \vec{M_{pp}}$ in addition to the structural stiffness $\vec{K_{pp}}$. Note that all the components in $\vec{U_d}$ are assumed to be dependent, and we only need $\vec{U_p}$ to describe the mode shapes fully. Hence \cref{eq:6by6} can also be written as:
\begin{equation}\label{eq:DE}
\vec{D}\vec{U_p}=0,
\end{equation}
where $\vec{D}=\vec{K_{pp}}-\vec{K_{pd}}\vec{K_{dd}^{-1}}\vec{K_{dp}}$ is the system's overall dynamic stiffness matrix. After substituting the parameters, it's easy to prove that $\vec{D}$ is Hermitian, thus having only real eigenvalues. For any given wavenumber and incident angle, the first three mode frequencies of the reduced order model can be obtained by solving the characteristic equation $\begin{vmatrix}\vec{D}\end{vmatrix}=0$. Plotting the mode frequencies with respect to wavenumber $Q_x$ and $Q_y$ yields the 3D band structure of oblique propagating waves. The eigen-vectors associated with these eigen-frequencies represent the mode shapes in the reduced order model. 

\section{Result and Discussion}

The asymmetric nature of this unit cell has created a coupling between P and SV modes that leads to quite interesting phenomena. While it may appear that the resonators only interact with P waves, the induced moment from the motion of the head interacts with the shear wave in the cell walls. The resulting band structure, therefore, demonstrates inherent and fundamental mode mixing. 

\begin{figure}[!ht]
	\begin{subfigure}[b]{0.5\linewidth}
		\centering\includegraphics[height=170pt]{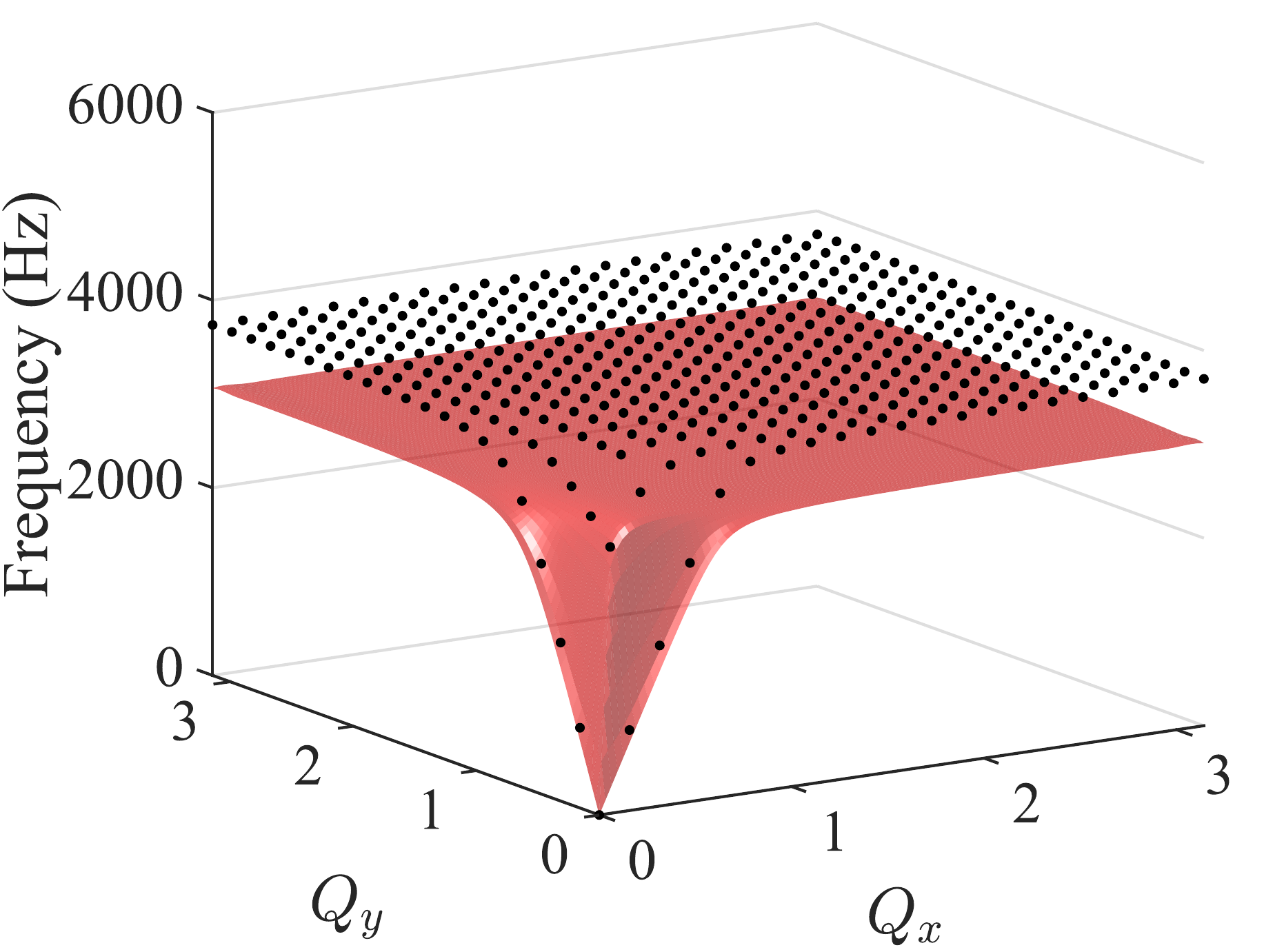}
		\caption{\label{fig:f1compare}}
	\end{subfigure}%
	\begin{subfigure}[b]{0.5\linewidth}
		\centering\includegraphics[height=170pt]{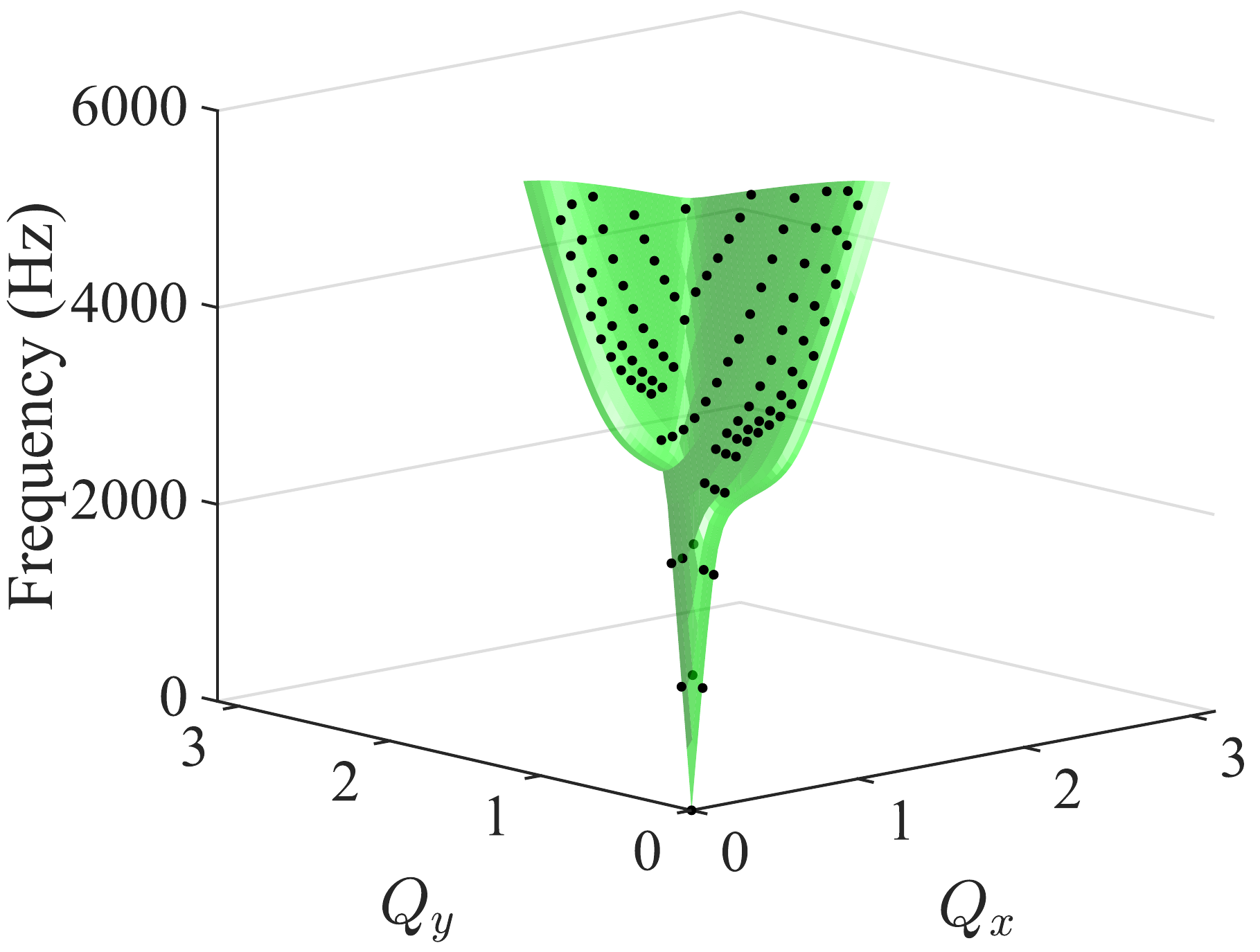}
		\caption{\label{fig:f2compare}}
	\end{subfigure}
	\begin{subfigure}[b]{0.5\linewidth}
		\centering\includegraphics[height=170pt]{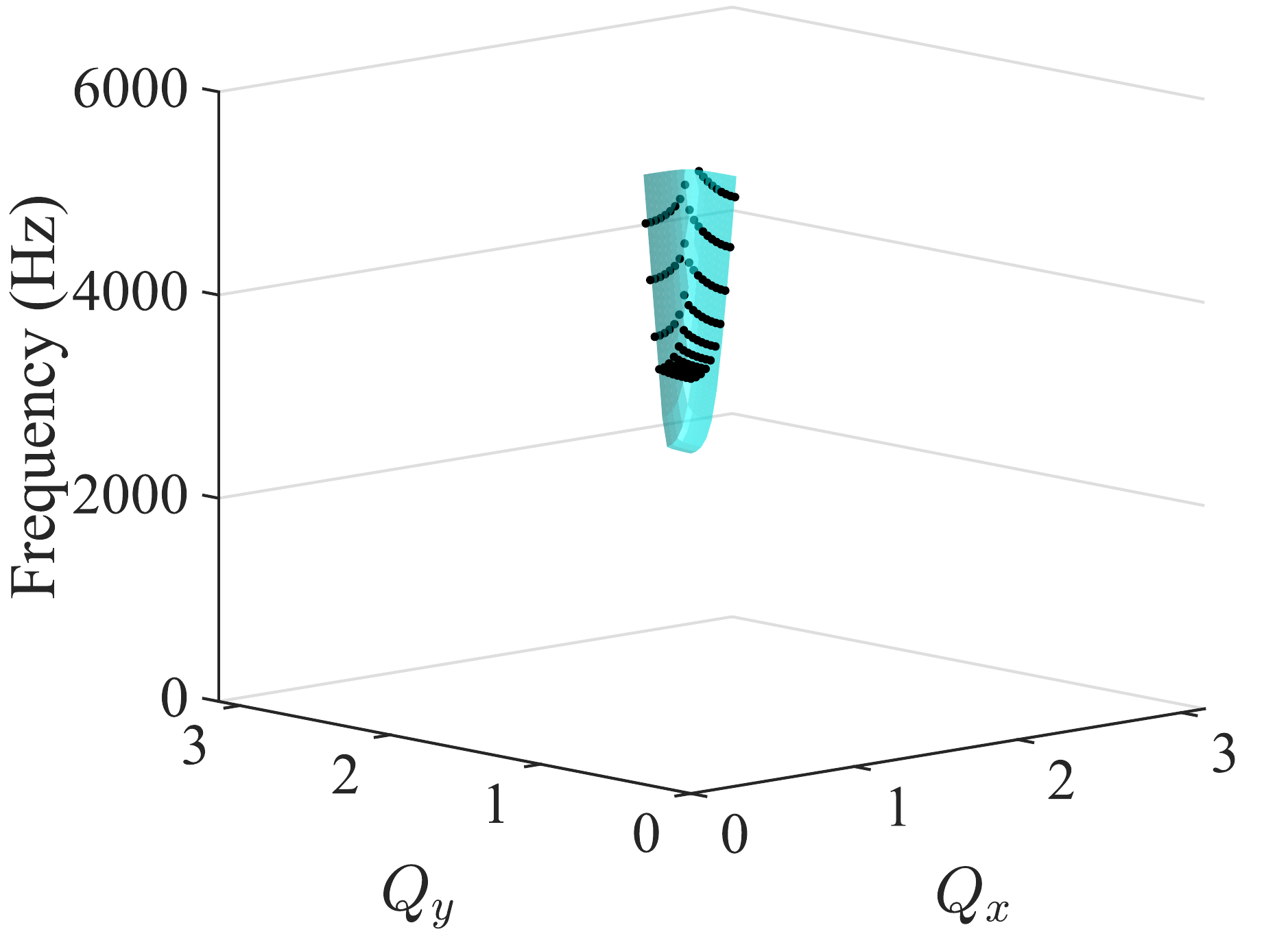}
		\caption{\label{fig:f3compare}}
	\end{subfigure}%
	\caption{\label{fig:3dcompares} Comparison of the 3D band structure under 6000 Hz. The colored surfaces are results from numerical method. The dots are obtained from analytical approach. (\subref{fig:f1compare})~through (\subref{fig:f3compare}) show the first, second and third modes, in order.} 
\end{figure}

\begin{figure}[!ht]
	\centering\includegraphics[height=180pt]{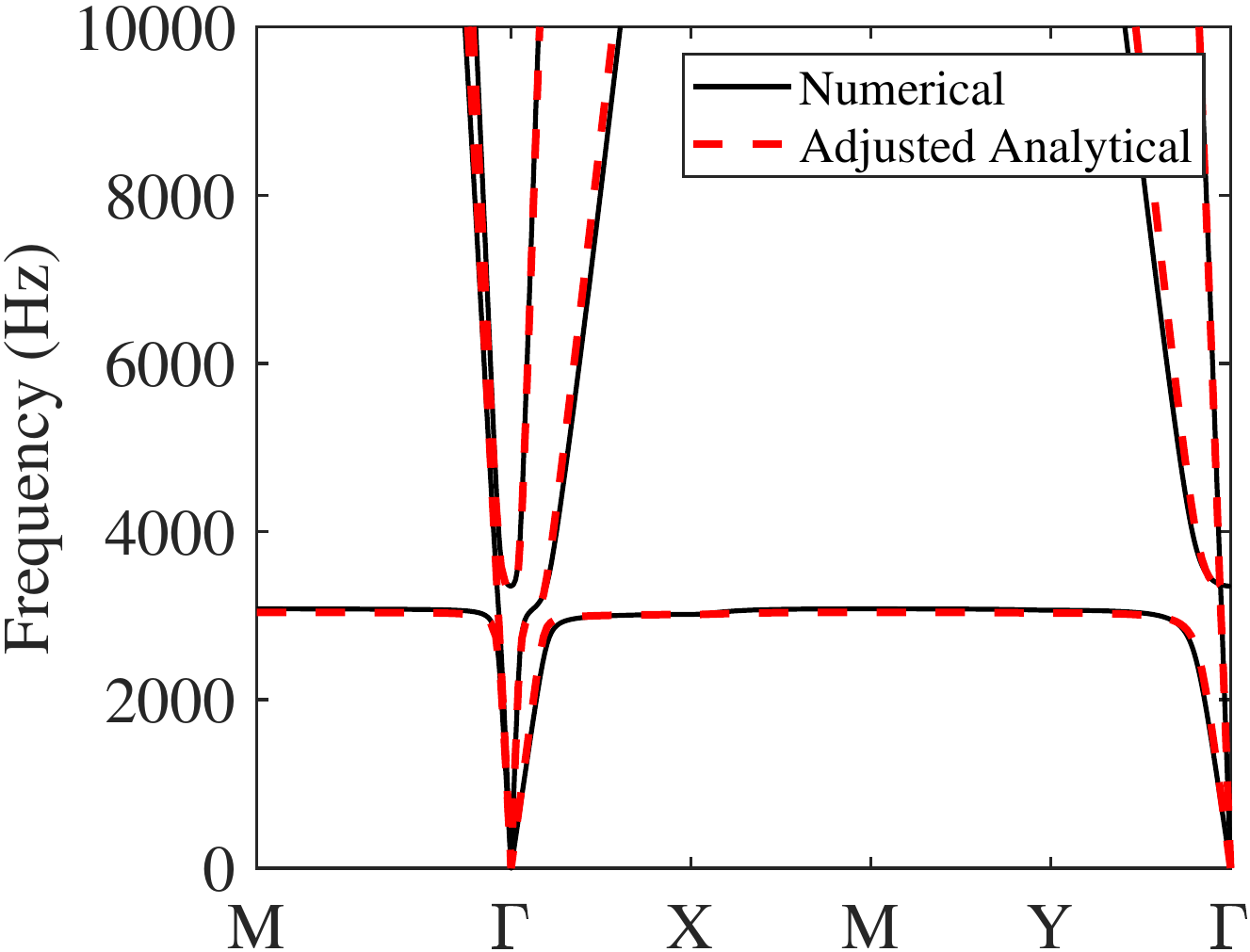}
	\caption{\label{fig:ibzcorr} Comparison of dispersion curves along paths connecting high symmetry $\vec{k}$ points from FE simulations (\cref{fig:bzo}) and based on the 3 DOF reduced order approach with adjusted $\beta_i$.}
\end{figure}

\begin{figure}[!ht]
\begin{subfigure}[b]{0.5\linewidth}
\centering\includegraphics[height=140pt]{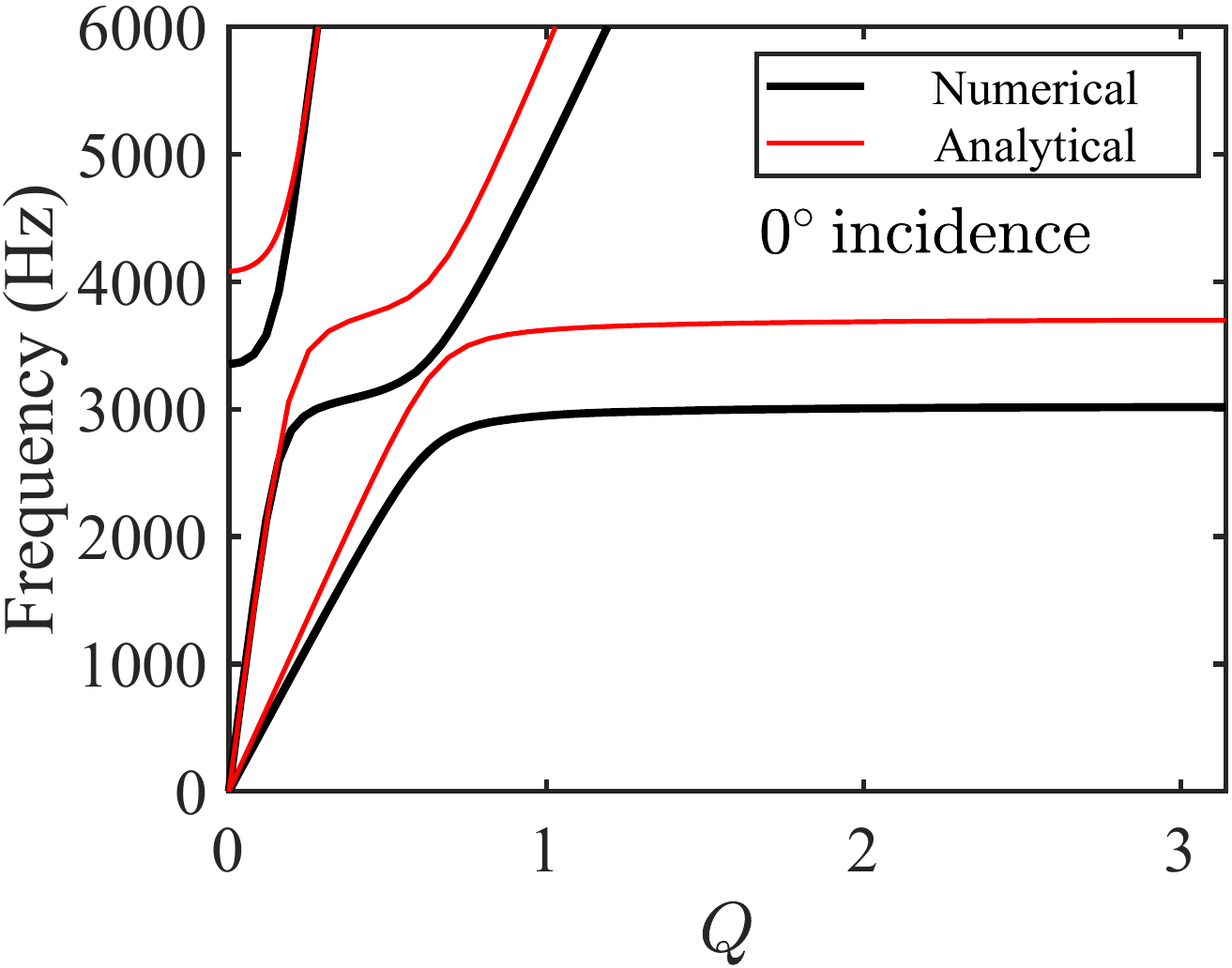}
\caption{\label{fig:dc0compare}}
\end{subfigure}%
\begin{subfigure}[b]{0.5\linewidth}
\centering\includegraphics[height=140pt]{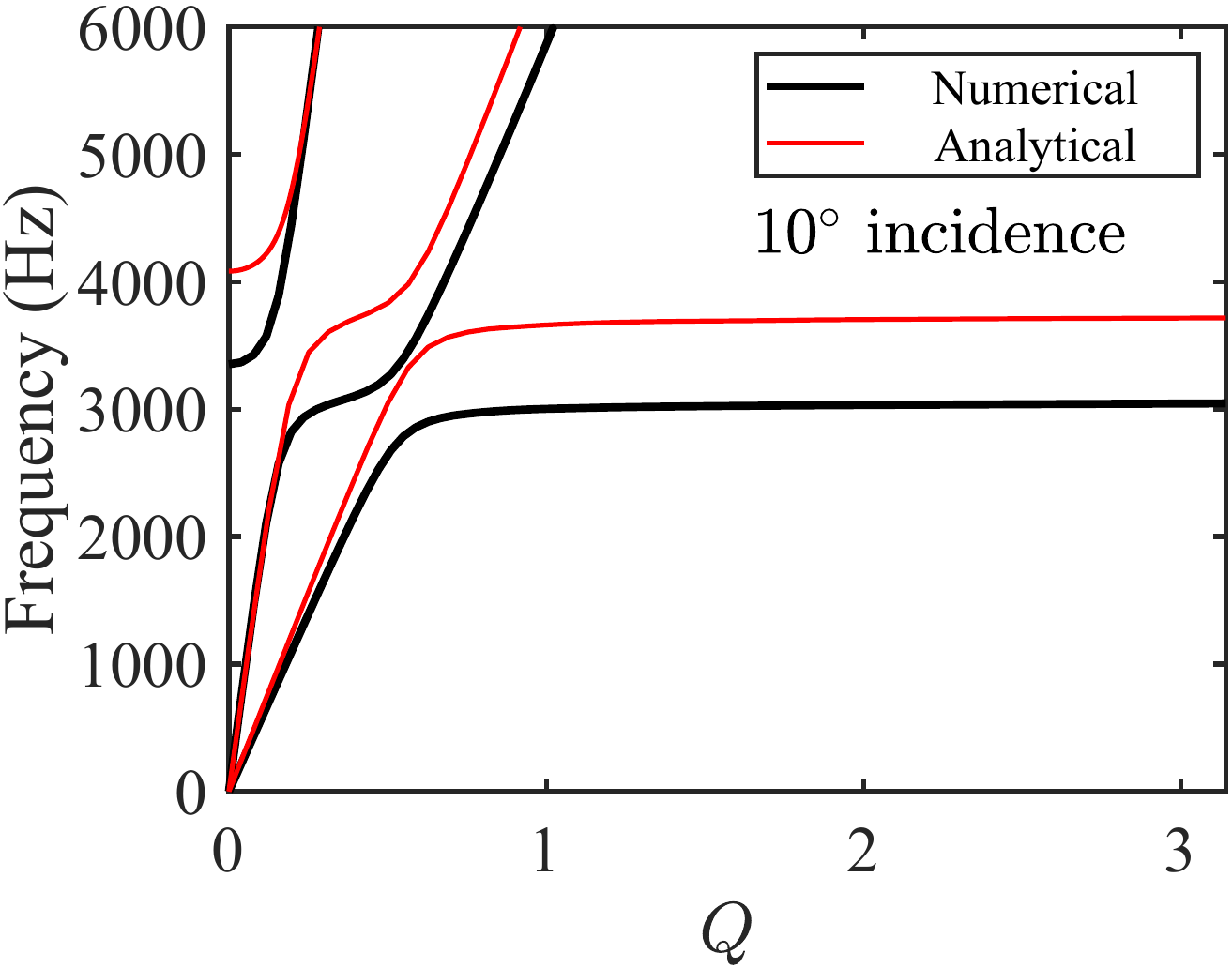}
\caption{\label{fig:dc10compare}}
\end{subfigure}\\
\begin{subfigure}[b]{0.5\linewidth}
\centering\includegraphics[height=140pt]{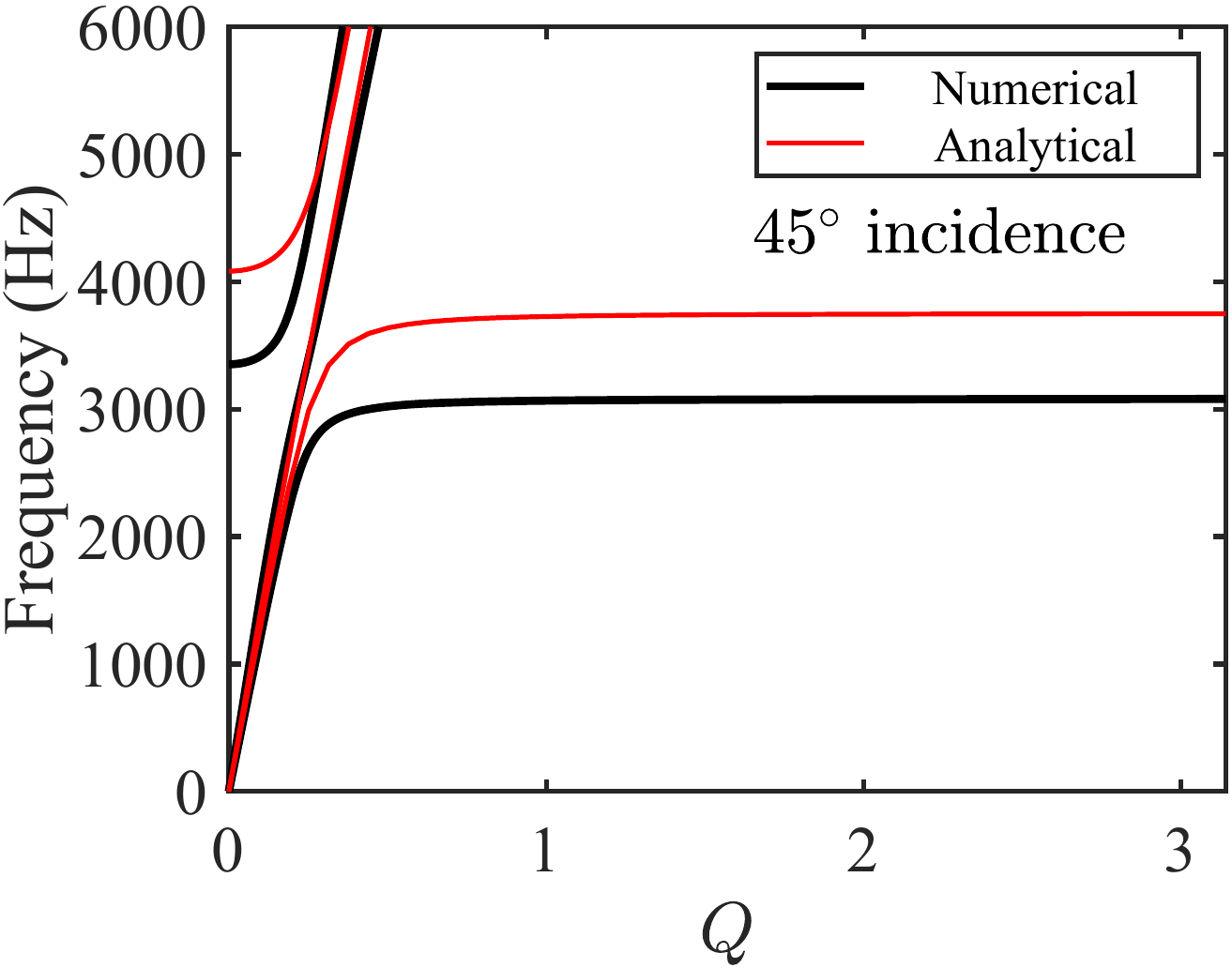}
\caption{\label{fig:dc45compare}}
\end{subfigure}%
\begin{subfigure}[b]{0.5\linewidth}
\centering\includegraphics[height=140pt]{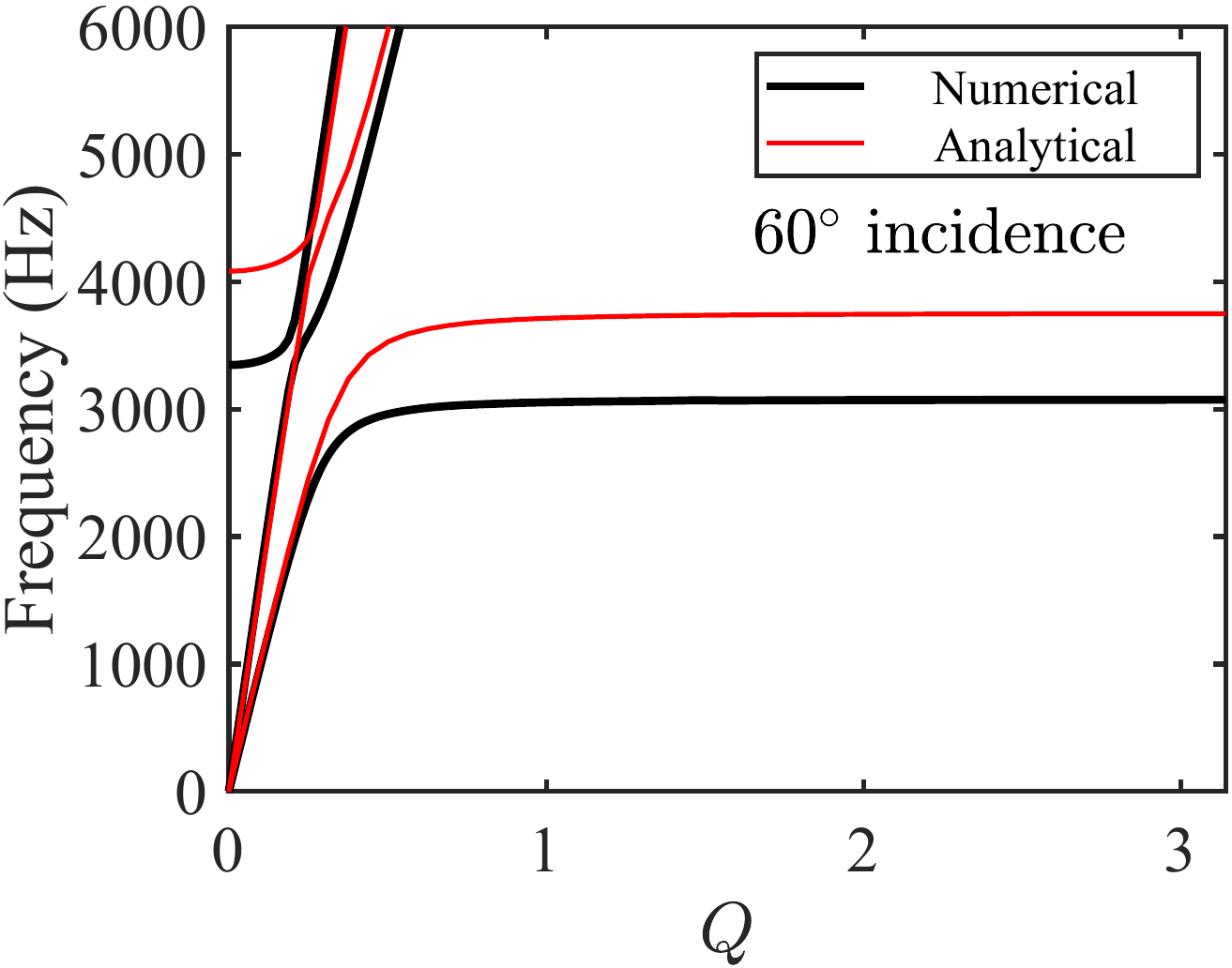}
\caption{\label{fig:dc60compare}}
\end{subfigure}\\
\begin{subfigure}[b]{0.5\linewidth}
\centering\includegraphics[height=140pt]{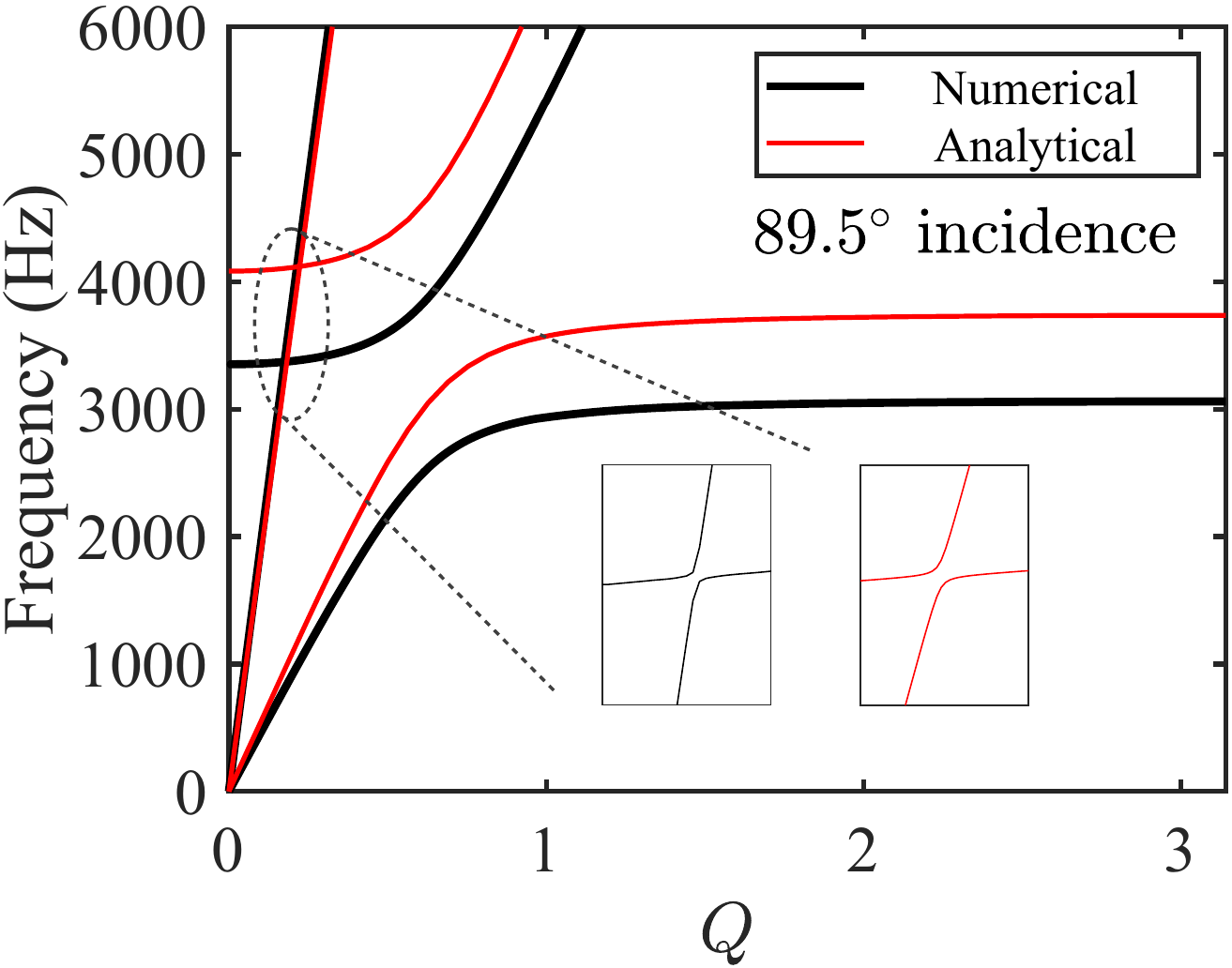}
\caption{\label{fig:dc895compare}}
\end{subfigure}%
\begin{subfigure}[b]{0.5\linewidth}
\centering\includegraphics[height=140pt]{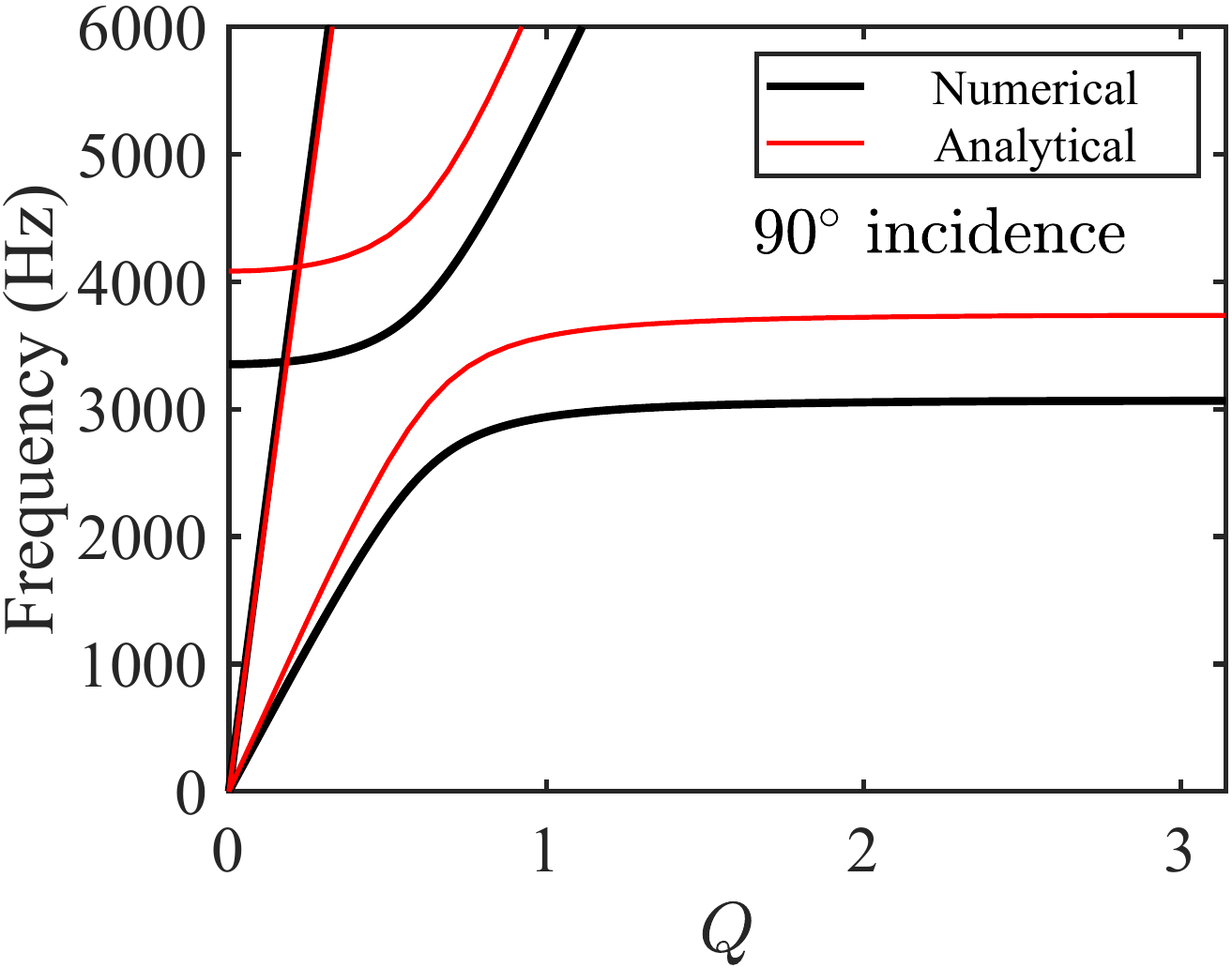}
\caption{\label{fig:dc90compare4dots}}
\end{subfigure}\\
\caption{\label{fig:dccompare} Partial band structure comparison. The red curves represent analytical results while the black ones are numerical results. (\subref{fig:dc0compare})~through (\subref{fig:dc90compare4dots}) are the band structures for 0$^\circ$, 10$^\circ$, 45$^\circ$, 60$^\circ$, 89.5$^\circ$, and 90$^\circ$ incidences, in order.} 
\end{figure}

The first 3 modes of the band structure for oblique in-plane waves are shown in \cref{fig:3dcompares} and comparison is made between the numerical COMSOL results and those of the reduced order analytical approach. Using the symmetry, the band structure is plotted only in the region where $Q_x$ and $Q_y$ are positive and vary from $0$ to $\pi$. Since the frequency band gap is expected to be located at around 3 kHz, the frequency range of interest will be $0$ to 6 kHz. Generally the results from analytical method agree well with the band structure from numerical approach. FE results are shown as colored surfaces while the dots represent the mode frequencies calculated using the analytical method. \Crefrange{fig:f1compare}{fig:f3compare} show the first, second and third modes, respectively.
\Cref{fig:f1compare} shows the acoustic branch of oblique SV wave. \Cref{fig:f2compare} shows the mixed branch of P and SV wave modes. \Cref{fig:f3compare} shows the optical branch of oblique P wave. In the quasi-static limit, as wavenumber $Q$ approaches the origin, the slopes of P wave and SV wave acoustic branches are proportional to the wave phase velocities in the media, and they are well approximated by the results from analytical calculation. However, it can be clearly observed that the entire calculated band gap is about 700 Hz higher than expected. The reason for such discrepancy is that the calculated resonator stiffness $\beta_i$ is overestimated by assuming part of the structure as rigid, thus rendering the resonance frequency higher than numerical simulation result. Furthermore, the rotational moment of inertia, particular at the resonator head is neglected. Including this rotational DOF at the resonator head should improve the results, though. While easily doable, the 4 DOF analysis is more convoluted without adding major clarity to the physical understanding of the problem. Instead, one can introduce a correction by recalculating $\beta_i$ in such a way that the resonance frequency of an isolated discrete resonator, $\omega_i = \sqrt{\beta_i/m_i}$, matches any estimate of the actual isolated resonator, i.e. with fixed boundary conditions at the root. With this correction, the band structure along paths connecting the corners of the irreducible Brillouin zone was calculated. \Cref{fig:ibzcorr} shows the results of this calculation, which turn out to be very close with FE results near resonances (reproduced from \cref{fig:bzo}). In the remainder of this paper, however, we have not used this correction to maintain consistency. 
\Crefrange{fig:dc0compare}{fig:dc90compare4dots} show partial dispersion diagrams for 0$^\circ$, 10$^\circ$, 45$^\circ$, 60$^\circ$, 89.5$^\circ$, and 90$^\circ$ propagation directions in order. 

The pattern of transition in dispersion curves can be seen in \cref{fig:dccompare}. The slopes of the two branches at the origin are related to wave velocities. The P wave velocity in the metamaterial reaches its minimum while SV wave velocity reaches its maximum when the waves are propagating at 45$^\circ$. As shown in \cref{fig:dc45compare}, the mixed branch of P and SV wave modes becomes almost a straight line. As the angle increases, P wave velocity increases after 45$^\circ$ while SV wave velocity decreases. 

As the angle increase from 60$^\circ$ towards 90$^\circ$, avoided crossing, or level repulsion in band structure becomes more evident.
The SV wave optical and P wave branches for 60$^\circ$ incidence almost intersect each other, as shown in \cref{fig:dc60compare}. However, the two branches exchange their slopes at the apparent crossing point and avoid each other due to level repulsion. If the numerical approach is performed on coarse wavenumber discretization \cite{Lu2018}, the avoided branches will be easily mistaken as real crossing branches. We note that the anti-crossing of dispersion curves can always be found until the propagation angle reaches exactly 90$^\circ$. For instance, at 89.5$^\circ$ the dispersion curves seem to have a real crossing point in \cref{fig:dc895compare}. However, zooming in the vicinity of the apparent crossing point indicates the opposite, as shown in the bottom right corner in \cref{fig:dc895compare}. \Cref{fig:dc895compare} implies mode anti-crossing instead of true intersection at 89.5$^\circ$ incidence, based on both numerical and reduced order analytical approaches. The reason for such avoided crossing is the small but non-negligible coupling between P and SV modes. As the P wave propagates at 89.5$^\circ$, although it is almost in $y$ direction, there is still a wave component in $x$ direction that causes the local resonance. In comparison, P wave is propagating in pure $y$ direction at 90$^\circ$ incidence and no local resonance occurs within the frequency range of interest. P wave and SV wave become decoupled only at 90$^\circ$ incidence and P wave acoustic branch has a real crossing point with SV wave optical branch. This can be further confirmed by the dynamic matrix in \cref{eq:DE} from the analytical method. Only when propagation angle is exactly 90$^\circ$, \cref{eq:DE} takes the form:
 \begin{equation}
 \label{eq:90}
 \left[\begin{matrix}
 D_{11} & D_{12} &0\\
  D_{12} & D_{22} &0\\
   0 & 0 &D_{33}\\
 \end{matrix}\right]
 \left[\begin{matrix}
 u^c_0\\u^i_0\\ v^c_0\end{matrix}\right]=0.
 \end{equation}
In other words since $k_x = 0$ at 90$^\circ$ angle, the stiffness components $D_{13}, D_{23}, D_{31}$ and $D_{32}$ in \cref{eq:DE} also vanish, thus making the cell displacement in $y$ direction decoupled from the other two DOFs in $x$ direction. Therefore, the eigenvectors must have the form:
\begin{equation}\label{eq:90ev}
\left[\begin{matrix}
u^c_0\\u^i_0\\ v^c_0\end{matrix}\right]=\left[\begin{matrix}
C_1\\C_2\\ 0\end{matrix}\right]\ \text{or} \left[\begin{matrix}
0\\0\\ 1\end{matrix}\right],
\end{equation}
where $C_1$ and $C_2$ are non-zero displacement components. Therefore at 90$^\circ$, the eigenvectors associated with $v^c_0$ are completely decoupled with $u^c_0$ and $u^i_0$.
Such real crossing point at 90$^\circ$ incidence has two degenerate eigenvalues but one can still find two associated linearly independent eigenvectors. 
However once the propagation angle changes from exactly 90$^\circ$, the coupling terms $D_{13}, D_{23}, D_{31}$ and $D_{32}$ appear again. Close to 90$^\circ$, $D_{13}, D_{23}, D_{31}$ and $D_{32}$ have smaller values, and shaper bending occurs in the dispersion curves.

\begin{figure}[!ht]
	 \begin{subfigure}[b]{0.5\linewidth}
	 \centering\includegraphics[height=140pt]{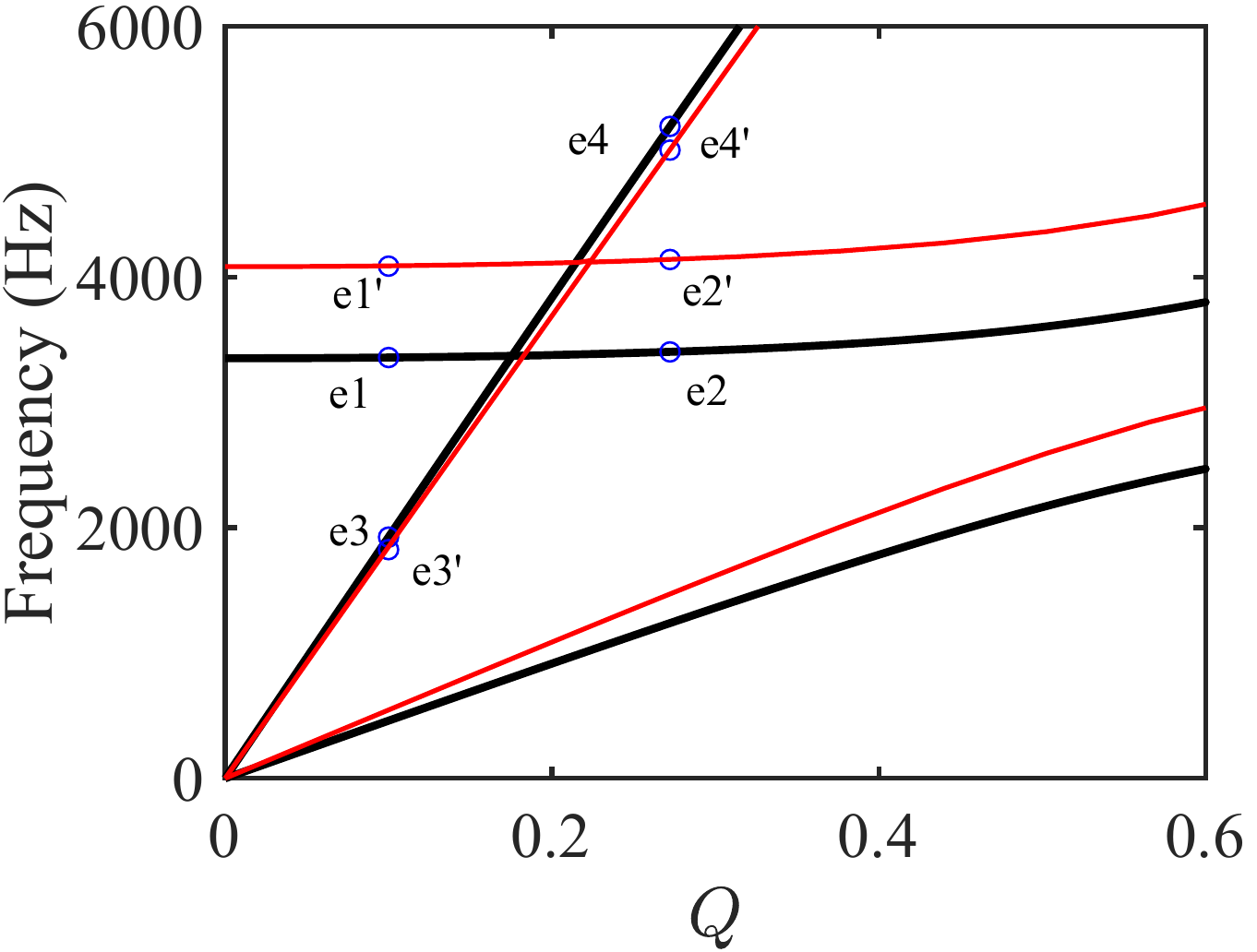}
	 \caption{\label{fig:dc895comsoldots}}
	 \end{subfigure}%
	\begin{subfigure}[b]{0.5\linewidth}
	\centering\includegraphics[height=140pt]{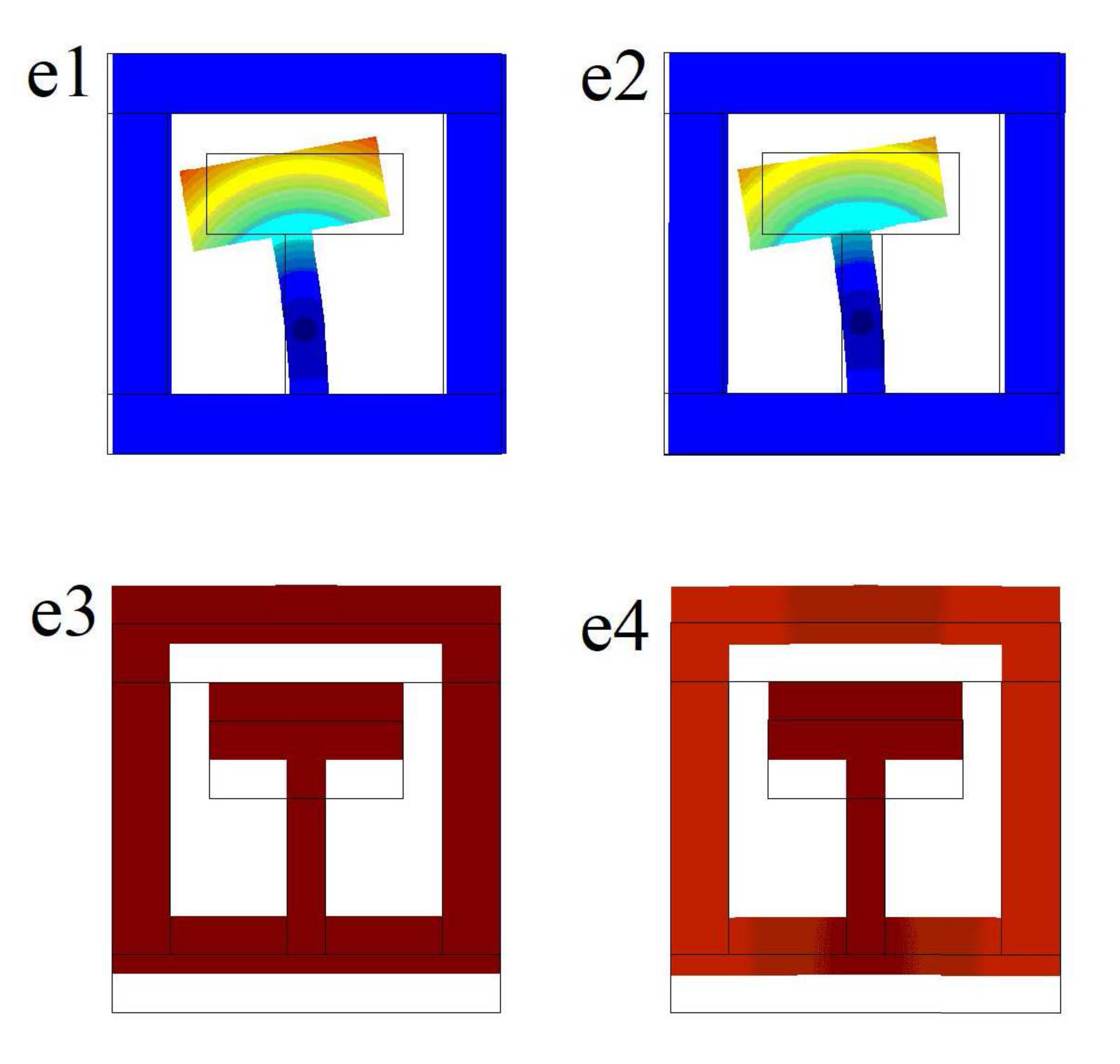}
	 \caption{\label{fig:emode}}
	 \end{subfigure}
	 \caption{\label{fig:895vicinity} (\subref{fig:dc895comsoldots}) The partial band structure in the vicinity of the apparent crossing point at propagation angle 89.5$^\circ$. The primed points e$n$' are associated with the reduced order analytical solution. (\subref{fig:emode}) The mode shapes at points e1, e2, e3, and e4 based on FE simulation results. The corresponding band structure and mode shape figures for 90$^\circ$ incidence are not shown here as they appear identical to (\subref{fig:dc895comsoldots}) and (\subref{fig:emode}), though with the important difference that the branches indeed cross in the real plane.} 
\end{figure}

The correct sorting of band structure is of prime importance as pointed out elsewhere\cite{Lu2018}. One way to identify different branches is zooming in the vicinity of apparent crossing point, as shown in \cref{fig:dc895compare}. However, if the band structure is calculated by finite element method, the zoom in method requires a very high level of resolution in wavenumber, which is computationally inefficient. Another technique proposed in literature\cite{Wu2004a,Yeh2016} uses the polarizations of modes as the criterion to distinguish real or apparent cross points. This method is based on the fact that the displacement fields change rapidly around the apparent crossing point. 
In \cref{fig:dc895comsoldots} we labeled four points e1, e2, e3, and e4. The corresponding mode shapes of these 4 points are obtained from finite element simulation and shown in \cref{fig:emode}.

Since these four points are chosen far away from the level repulsion area and are based on coarse discretization of wavenumber, it is very difficult to distinguish the difference between mode shapes at e1 and e2, or between e3 and e4, as shown in \cref{fig:emode}. The similar mode shapes may make one believe that points e1 and e2 belong in the same branch and points e3 and e4 are on the other presumably continuous branch, which are both incorrect conclusions. In other words, when the points are chosen not close enough to the apparent crossing point, the difference in their eigenvectors could be very small, and the simulated mode shapes could lead to incorrect sorting. To further demonstrate the changes in mode shapes around apparent crossing points, we picked the corresponding points e1', e2', e3', and e4' from the analytical result, as shown in \cref{fig:dc895comsoldots}, and calculated their eigenvectors, see \cref{tab:emodes}. The eigenvectors at 89.5$^\circ$ show that the displacement of the cell in $y$ direction, $v^c_0$, changes its sign between points e1' and e2', and the change in sign also occurs for the displacement of resonator in $x$ direction, $u^i_0$, between e3' and e4'. Note that we ensured to select eigenvectors, for which the dominant components of the the eigenvectors ($u^i_0$ in the former case and $v^c_0$ in the latter) are essentially unchanged, so the sign or phase changes in other components have physical meaning in terms of change in mode shape. The small amplitudes of the components that change signs compared to the dominant components make identifying such differences from the simulated mode shapes difficult. The third component of the 89.5$^\circ$ eigenvector is not significantly affected in the transition from e1' to e2', but in the case of transition from e3' to e4', it moves one quadrant in the complex domain (sign of the imaginary part changes). These sign flips (in 89.5$^\circ$ case) prove that the points e1 and e2 are not on the same branch and neither are e3 and e4, i.e. we are observing an avoided level crossing. Furthermore, for 89.5$^\circ$ incidence, all the eigenvalue components have non-zero value, which indicates coupling between displacement components, in contrast to the decoupled case of 90$^\circ$ as shown in \cref{eq:90ev}. These points are located in the 90$^\circ$ incidence band structure at the same $Q$ values with e1' to e4' and their eigenvectors are shown in \cref{tab:emodes}. The eigenvector components with zero values indicate decoupling of motions. The fact that there is no change of sign in eigenvectors also proves real branch crossing at 90$^\circ$ incidence.

\begin{table}
\caption{\label{tab:emodes} Calculated displacement eigenvectors for 89.5$^\circ$ and $90$ incidences.}
\begin{ruledtabular}
\begin{tabular}{*2{C{0.5in}} *4{C{1.2in}}}
 & &e1'&e2'&e3'&e4'\\ 
\hline
\multirow{3}{*}{89.5$^\circ$}&$u^c_0$&5.04e-2-i1.76e-1&1.37e-1-i1.59e-1&7.91e-4+i1.45e-4&1.37e-3-i6.36e-4\\
&$u^i_0$&-2.58e-1+i9.49e-1&-6.09e-1+i7.64e-1&1.05e-3-i7.31e-4&-1.63e-3+i3.08e-3\\
&$v^c_0$&1.36e-4&-8.43e-4&9.99e-1&9.99e-1\\
\hline
\multirow{3}{*}{90$^\circ$}&$u^c_0$&1.80e-1+i4.80e-3&1.89e-1+i1.36e-2&0&0\\
&$u^i_0$&-9.84e-1&-9.81e-1&0&0\\
&$v^c_0$&0&0&1&1\\
\end{tabular}
\end{ruledtabular}
\end{table}

A similar analysis may be performed on the mode shapes or eigenvectors calculated in the FE calculations. To demonstrate this, one can define aggregate cell and inclusion displacements by averaging the point-wise quantities over the associated domains. Then, normalized displacement eigenvector components may be defined as
\begin{align}
	d^1_j &= \bar{u}^c_j/\bar{v}^c_j, \label{eq:d1}\\
	d^2_j &= \bar{u}^i_j/\bar{v}^c_j, \\
	d^3_j &= \bar{u}^c_j/\bar{u}^i_j,
\end{align}
where $\ \bar{ }\ $ denotes the average numerical value, and $j$ represents the mode number in the order of increasing frequency. Note that these three values are obviously inter-dependent and are depicted here only for clarity of presentation. Further, define amplitude and argument of these quantities as
\begin{equation}
	d^k_j=r^k_je^{i\phi^k_j}. \label{eq:drphi} 
\end{equation}

\begin{figure}[!ht]
	\begin{subfigure}[b]{0.5\linewidth}
		\centering\includegraphics[height=150pt]{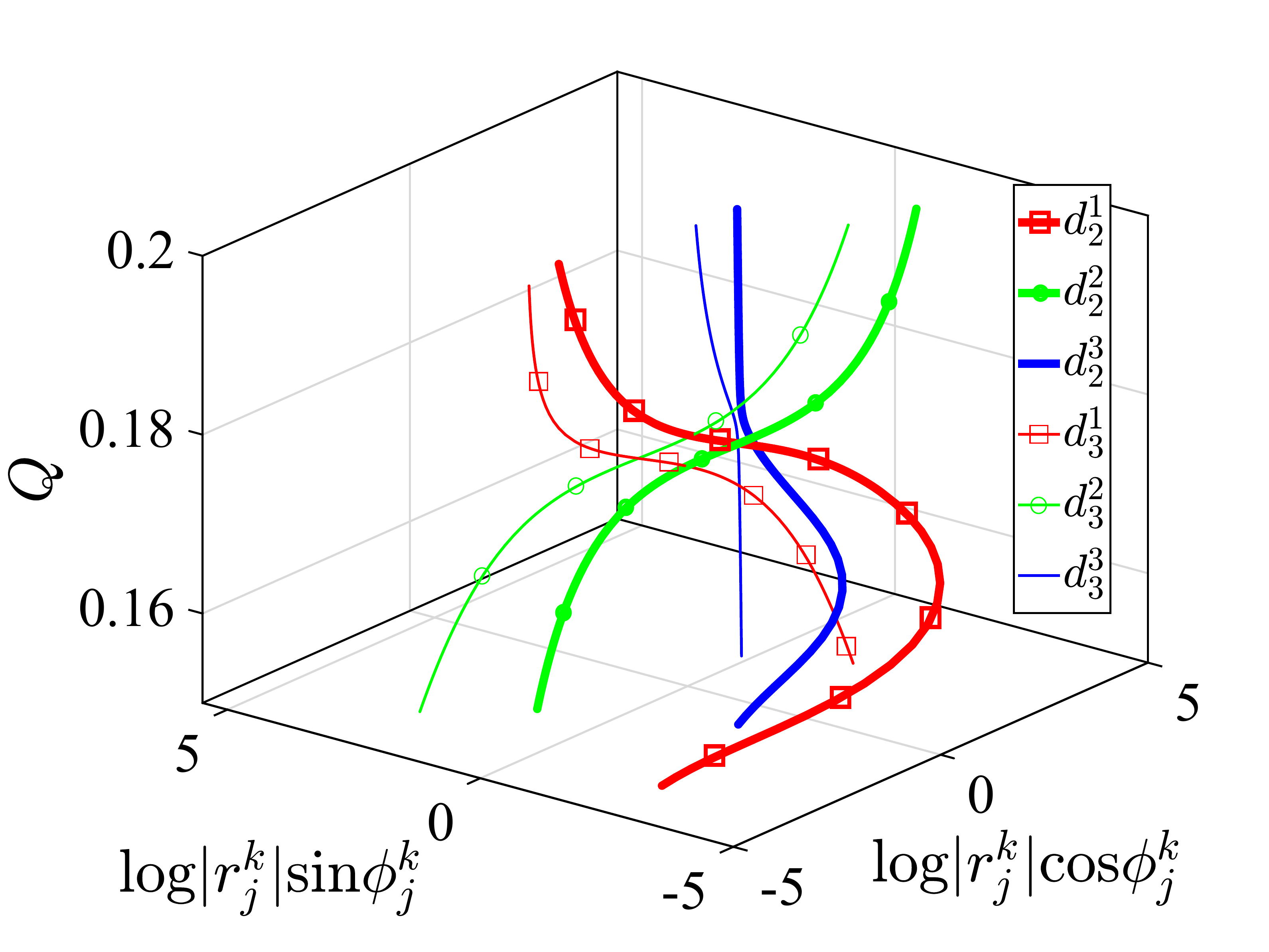}
		\caption{$\theta = 85^\circ$\label{fig:85a2}}
	\end{subfigure}%
	\begin{subfigure}[b]{0.5\linewidth}
		\centering\includegraphics[height=150pt]{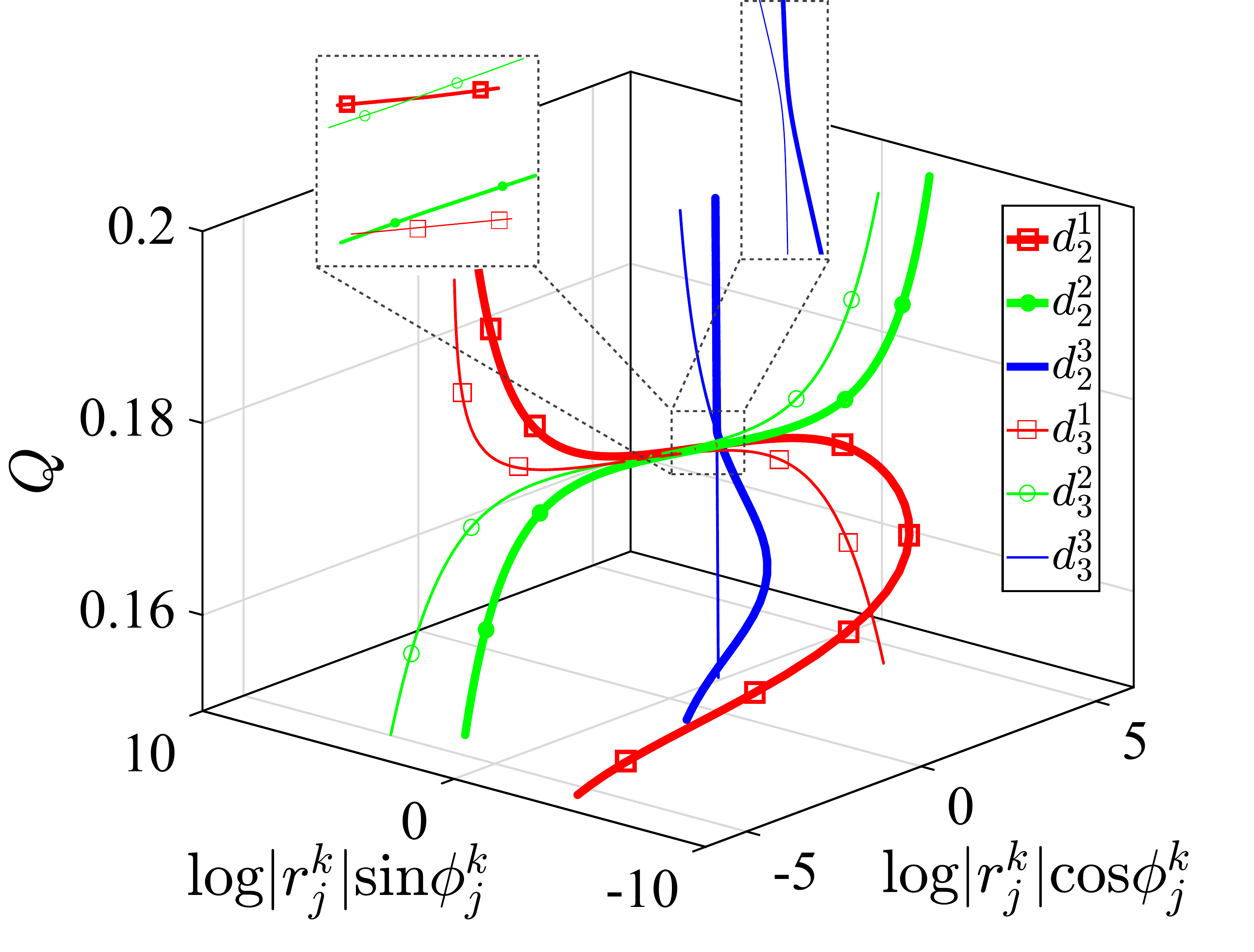}
		\caption{$\theta = 89.5^\circ$\label{fig:89a2}}
	\end{subfigure}\\
	\begin{subfigure}[b]{0.5\linewidth}
		\centering\includegraphics[height=150pt]{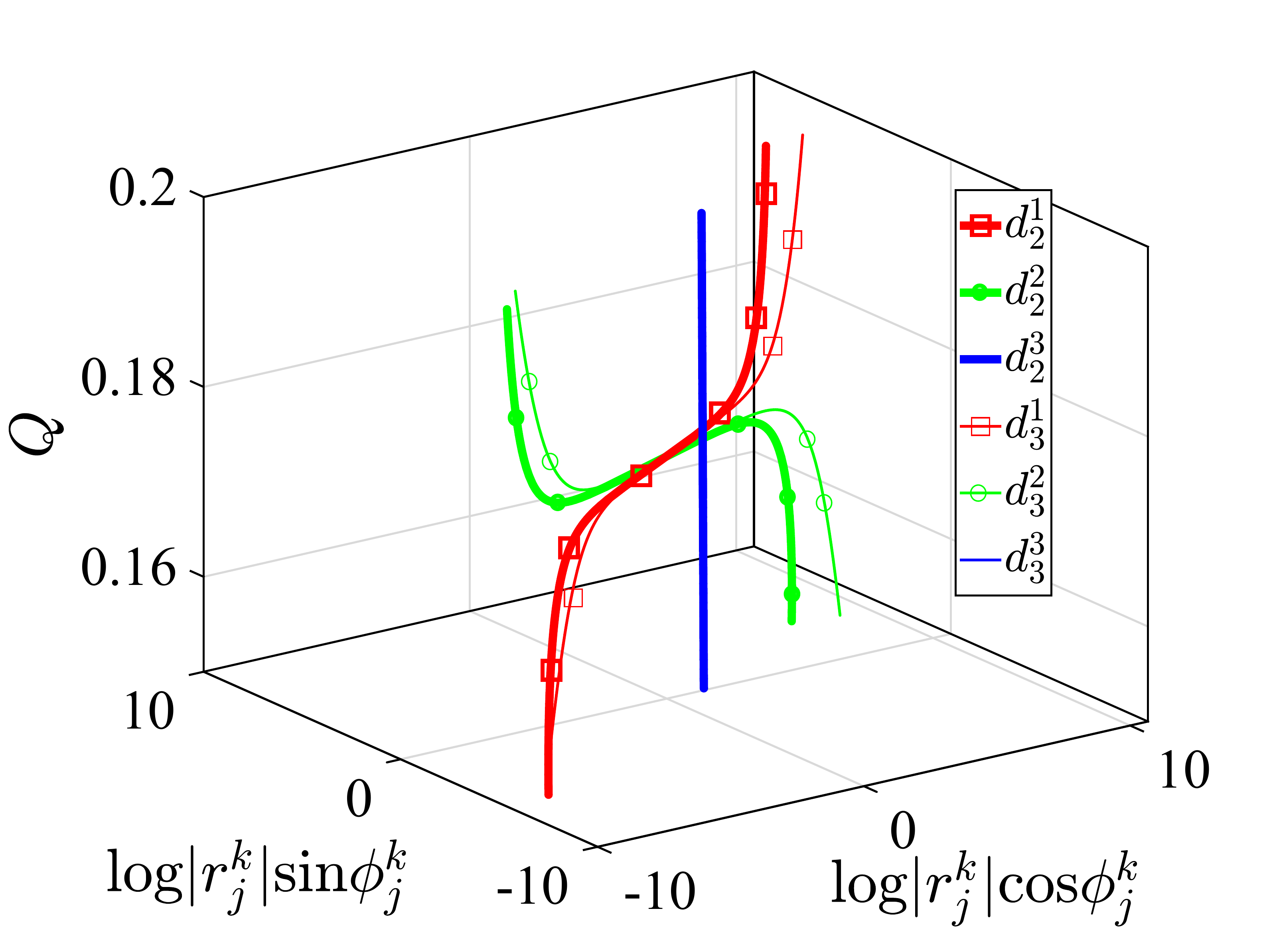}
		\caption{$\theta = 90^\circ$\label{fig:90a2}}
	\end{subfigure}%
	\begin{subfigure}[b]{0.5\linewidth}
		\centering\includegraphics[height=150pt]{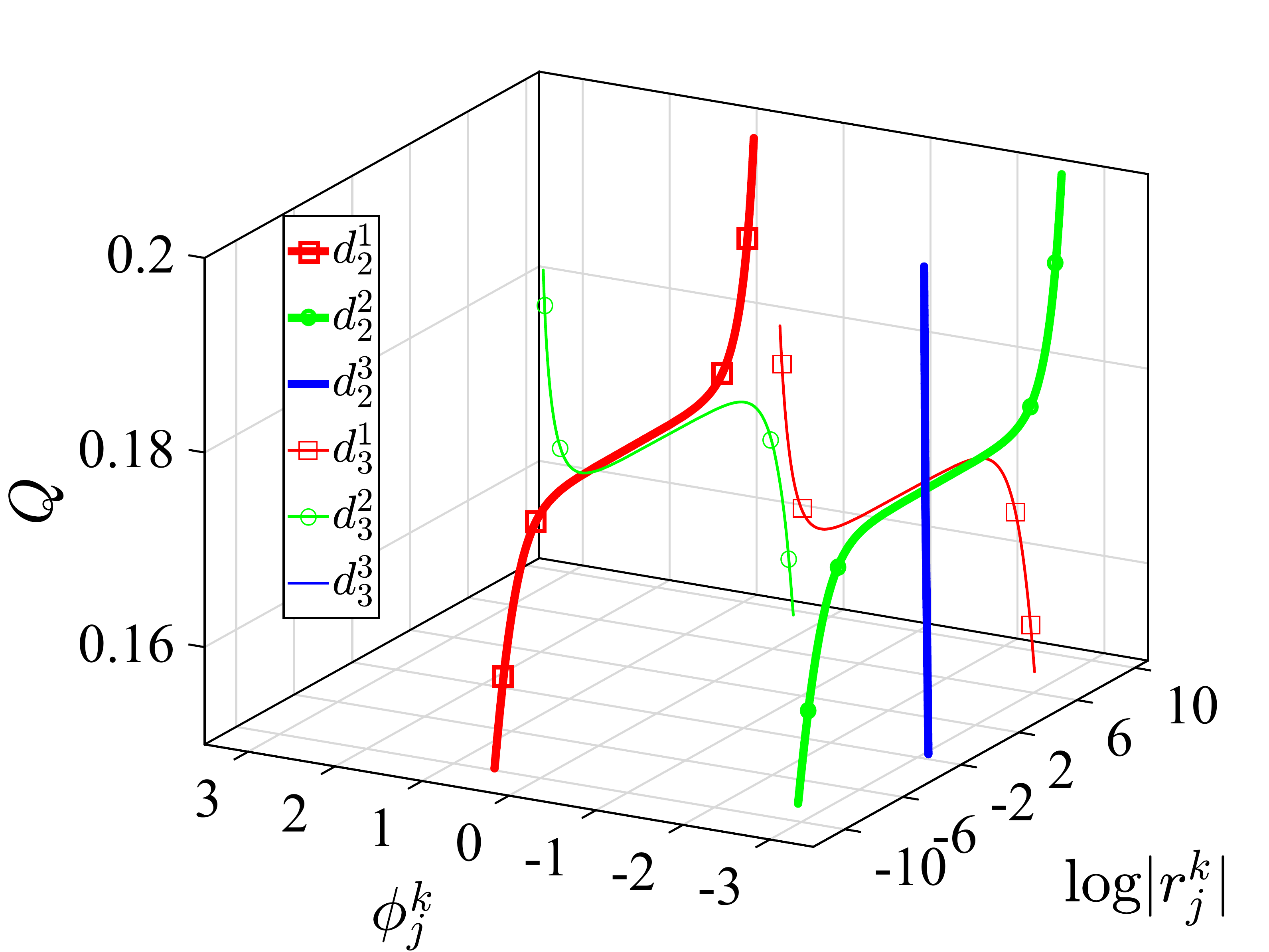}
		\caption{$\theta = 90^\circ$\label{fig:90a1}}
	\end{subfigure}
	\caption{\label{fig:89lr} Representation of mode shape behavior near apparent ($\theta = 85^\circ,\ 89.5^\circ$) and real ($\theta = 90^\circ$) crossings based on quantities defined in \crefrange{eq:d1}{eq:drphi}. The high resolution insets in (\subref{fig:89a2}) show that the mode shapes do not intersect. The different representation in (\subref{fig:90a1}) depicts the difference between the modes in the real crossing at $\theta = 90^\circ$ more clearly as the transformation used in (\subref{fig:90a2}) artificially connects the neighborhoods $r \rightarrow 0$ and $r \rightarrow \infty$ through a $\pi$ change in phase.} 
\end{figure}

Results are plotted in \cref{fig:89lr} for $\theta = 85^\circ$, $89.5^\circ$, $90^\circ$ propagation angles from $Q=0.15$ to $Q=0.2$, where the second and third mode branches ($j=2,\ 3$) have apparent (85$^\circ$ and 89.5$^\circ$) or real ($90^\circ$) crossings. The amplitudes $r_j^k$ could become very big or very small, which makes it difficult to really distinguish mode shapes that may have very different phases. The transformed trajectories shown in \cref{fig:89lr} allows one to separate them. Nevertheless, near the apparent or real crossings, one would still need relatively high resolution to distinguish the modes. In other words, the mode shape analysis presented here does not resolve the issues that prompted the sorting algorithm proposed by Lu and Srivastava\cite{Lu2018}. The case of $\theta = 90^\circ$ requires further analysis for two reasons. First, regardless of the resolution, the trajectories of the two modes $j = 2,\ 3$ appear to intersect. Furthermore, the amplitudes become extremely small (see amplitudes in \cref{fig:90a1}), and the phases appear to be very close as well (when $r \rightarrow 0$ an apparent $\pi$ jump in phase as seen in \cref{fig:90a1} does not indicate discontinuity). It is fair to say that in these cases $d^k_j$ for $k = 1,\ 2$ is essentially zero and the mode shape is a pure $d^3_j = \bar{u}^c_j / \bar{u}^i_j \approx const$. In other words, at the exact intersection of the two branches, the eigenvalues of two very different eigenvectors coalesce at one point. Therefore, in order to continuously follow the same physical modes (i.e. the one with very small $r^1_j$ and $r^2_j$ versus the one with very large values of these quantities) across this point, one has to switch from $j = 2$ branch to $j = 3$ branch and vice versa, as also seen in \cref{fig:90a1} by following the paths when $\log r^k_j \rightarrow -\infty$ by an inconsequential $\Delta \varphi = \pi$ shift in phase (as it is also the case, analytically, when $\log r^k_j \rightarrow \infty$). 

\section{Summary and Conclusions}

This paper presents a numerical and analytical study of the band structure of an acoustic metamaterial for oblique in-plane waves.
An analytical method with reduced order model is established to approximate the band structure. Dispersion curves and mode shapes for different incident angles are shown. The studied metamaterial exhibits mixed branch of P and SV wave modes except when the propagation angle is exactly 90$^\circ$.
We also report on the occurrence of avoided level crossings, which are related to the existence of exceptional points in complex domain\cite{Lu2018}. With relatively high computational efficiency, the analytical method can calculate partial band structure which agrees well with numerical results, particularly when adjusted for the exact resonance frequency of a single discrete resonator (with rigid boundaries and not connected to the cell) and can provide support in mode identification and band sorting. With proper adjustments in parameters, this analytical method will be applicable for other metamaterials that have similar unit cell structure.
\bibliographystyle{aipnum4-1}
\bibliography{Manuscript}

\end{document}